\documentclass[epsf,useAMS,usenatbib, usegraphicx]{mn2e}
\usepackage{graphicx}
\usepackage{epsfig}
\usepackage{amsmath}
\usepackage{amssymb}
\usepackage{rotating}
\usepackage{color}
\usepackage{pifont}
\usepackage{longtable}
\usepackage{pdflscape,lscape}
\usepackage{float}
\usepackage{placeins}

\newcommand{\teff}{$T_{\rm eff}$}
\newcommand{\logg}{$\log g$}
\newcommand{\vsini}{$v \sin i$}
\newcommand{\kms}{km\,s$^{-1}$}
\newcommand{\ds}{$\delta$\,Scuti}
\newcommand{\vr}{$v_{r}$}
\newcommand{\ppuls}{$P_{\rm puls}$}
\newcommand{\porb}{$P_{\rm orb}$}
\newcommand{\tzero}{$T_0$}
\newcommand{\tess}{{\it TESS}}

\usepackage{booktabs}
\usepackage{upgreek}
\usepackage{url}

\usepackage{natbib}
\usepackage[bookmarks=false]{hyperref}
\hypersetup{colorlinks = true, linkcolor = green, anchorcolor = red, citecolor = blue, filecolor = red, pagecolor = red, urlcolor = red}
 \UseRawInputEncoding
\title[Comprehensive study of AI\,Hya]{Comprehensive spectroscopic and photometric study of pulsating eclipsing binary star AI\,Hya}

\author[F. Kahraman Ali\c{c}avu\c{s} et. al.]{F. Kahraman Ali\c{c}avu\c{s}$^{1,2}$\thanks{E-mail: filizkahraman01@gmail.com}, T. Pawar$^{3}$\thanks{E-mail: pawar@ncac.torun.pl}, K. G. He\l{}miniak$^{3}$, G. Handler$^{4}$, A. Moharana$^{3}$,  \and F. Ali\c{c}avu\c{s}$^{1,2}$,  P. De Cat$^{5}$, F. Leone$^{6,7}$, G. Catanzaro$^{7}$, M. Giarrusso$^{7,8}$, N. Ukita$^{9,10}$, \and E. Kambe$^{11}$
\\
$^{1}$\c{C}anakkale Onsekiz Mart University, Faculty of Science, Physics Department, 17100, Canakkale, Turkey\\
$^{2}$\c{C}anakkale Onsekiz Mart University, Astrophysics Research Center and Ulup{\i}nar Observatory, TR-17100, Çanakkale, Turkey\\
$^{3}$Nicolaus Copernicus Astronomical Center, Department of Astrophysics, ul. Rabia\'{n}ska 8, PL-87-100 Toru\'{n}, Poland\\
$^{4}$Nicolaus Copernicus Astronomical Center, Polish Academy of Sciences, Bartycka 18, PL-00-716 Warsaw, Poland\\ 
$^{5}$Royal Observatory of Belgium, Ringlaan 3, B-1180 Brussel, Belgium\\
$^{6}$Dipartimento di Fisica e Astronomia, Sezione Astrofisica, Universit?a di Catania, Via S. Sofia 78, I-95123 Catania, Italy\\
$^{7}$INAF, Osservatorio Astrofisico di Catania, Via S. Sofia 78, I-95123 Catania, Italy\\
$^{8}$University of Florence, Department of Physics and Astronomy, Via Giovanni Sansone 1, I-50019 Sesto Fiorentino, Italy\\
$^{9}$Okayama Astrophysical Observatory, National Astronomical Observatory of Japan, 3037-5 Honjo, Kamogata, Asakuchi, Okayama 719-0232, Japan\\
$^{10}$The Graduate University for Advanced Studies, 2-21-1 Osawa, Mitaka, Tokyo 181-8588, Japan\\
$^{11}$Subaru Telescope, National Astronomical Observatory of Japan, 650 North Aohoku Place, Hilo, HI 96720, USA\\
}
\begin{document}

\date{Accepted ... Received ...; in original form ...}

\pagerange{\pageref{firstpage}--\pageref{lastpage}} \pubyear{2021}

\maketitle

\label{firstpage}

\begin{abstract}
 
The pulsating eclipsing binaries are remarkable systems that provide an opportunity to probe the stellar interior and to determine the fundamental stellar parameters precisely. 
Especially the detached eclipsing binary systems with (a) pulsating component(s) are significant objects to understand the nature of the oscillations since the binary effects in these systems are negligible. 
Recent studies based on space data have shown that the pulsation mechanisms of some oscillating stars are not completely understood.
Hence, comprehensive studies of a number of pulsating stars within detached eclipsing binaries are important.
In this study, we present a detailed analysis of the pulsating detached eclipsing binary system AI\,Hya which was studied by two independent groups with different methods. 
We carried out a spectroscopic survey to estimate the orbital parameters via radial velocity measurements and the atmospheric parameters of each binary component using the composite and/or disentangled spectra. 
We found that the more luminous component of the system is a massive, cool and chemically normal star while the hotter binary component is a slightly metal-rich object.
The fundamental parameters of AI\,Hya were determined by the analysis of binary variations and subsequently used in the evolutionary modelling. Consequently, we obtained the age of the system as 850\,$\pm$\,20\,Myr and found that both binary components are situated in the \ds\ instability strip. The frequency analysis revealed pulsation frequencies between the 5.5 -- 13.0\,d$^{-1}$ and we tried to estimate which binary component is the pulsating one. However, it turned out that those frequencies could originate from both binary components.

\end{abstract}


\section{Introduction}

To understand the universe, it is necessary to comprehend stars which are its building blocks. 
For a deep investigation of stars, we should know their basic stellar parameters such as mass ($M$), radius ($R$) and chemical composition. 
Binary stars, in particular the eclipsing ones, are the most suitable objects to derive these parameters as $M$ and $R$ can be derived with an accuracy better than 1\% \citep{2010A&ARv..18...67T, 2013A&A...557A.119S}. Therefore, these systems are substantial for a better understanding of the universe, our Galaxy, and, most directly, stellar evolution. However, eclipsing binary systems as such do not provide information about the stellar interior. This is where the pulsating stars come in. 
The oscillation frequencies of pulsating stars can be used to probe the stellar interior by applying asteroseismic methods, making eclipsing binary systems with (a) pulsating component(s) one of the most valuable tools to improve our knowledge of stellar evolution.

Various types of pulsating stars in different evolutionary states exist. 
Some of them, such as $\beta$\,Cephei, $\delta$\,Scuti, and $\gamma$\,Doradus stars \citep{2021Galax...9...28L, 2021Univ....7..369S}, are also found in eclipsing binary systems. 
The \ds\, variables are the most common pulsating stars found in eclipsing binaries because of their relatively short pulsation periods. The \ds\, stars are A to F-type dwarf or giant stars generally exhibiting pressure mode oscillations with periods between 18\,min and 8\,h and amplitudes below 0$^{m}$.1 in the V-band \citep{2010aste.book.....A}. Their theoretical instability strip (e.g. \citealt{2005A&A...435..927D}) indicates the location of objects in the Hertzsprung-Russell (H-R) diagram that are expected to show \ds-type oscillations. Thanks to space missions such as \textit{Kepler} \citep{2010Sci...327..977B} and the Transiting Exoplanet Survey Satellite \citep[\tess,][]{2014SPIE.9143E..20R}, we learned that \ds\, stars are also observed beyond the borders of the theoretical instability strip, showing the necessity to revise them \citep{2011A&A...534A.125U, 2014ApJ...796..118A, 2018MNRAS.476.3169B}. 
According to the latest catalog of \ds\ stars in eclipsing binaries, there are around 90 such objects \citep{2017MNRAS.470..915K}. 
This number is now increasing especially by the discoveries of new systems from the investigation of the space data \citep[e.g.][]{2022RAA....22h5003K, 2019A&A...630A.106G}. The pulsations of the \ds\, stars in eclipsing binaries are affected by the other binary component \citep{2017MNRAS.470..915K, 2017MNRAS.465.1181L}. Indeed, their pulsation period 
(\ppuls) 
decreases when the orbital period 
(\porb) 
becomes shorter and, hence, the other component approaches the pulsating component.
It was also thought that the tidal forces between the binary components can alter the pulsation axis \citep{2020MNRAS.494.5118K}.
The first observational proof of this was presented by \cite{2020NatAs...4..684H} thanks to the high-quality data of \tess. 
These authors showed that in some binary systems the pulsation axis can align with the orbital axis because of the tidal forces. This type of object is now known as tidally tilted pulsators and they are a clear proof of binary effects on pulsations.  

For a deep understanding of the effects of binarity on pulsations in eclipsing binary systems and on stellar evolution and structure, comprehensive investigations of such systems are necessary. AI\,Hya (V\,=\,9$^{m}$.35) is an eclipsing binary system with a \ds\, component consisting of a F2m and F0V star \citep{2015A&A...575A.117S}. 
It has an eccentric orbit and an orbital period of 8.289649(2) days \citep{2004AcA....54..207K}. Spectroscopic observations revealed that AI\,Hya is a double-lined binary system \citep{1988AJ.....95..190P}. In a recent study, an updated photometric analysis based on the \tess\ data of AI\,Hya was given which shows that the secondary component exhibits multiperiodic oscillations \citep{2020PASJ...72...37L}. However, no detailed spectral analysis with high-resolution spectra has been carried out for the system so far. Therefore, we provide a detailed photometric and spectral analysis of AI\,Hya in this study to reveal the true character of this interesting object. 

Two teams were working on this system independently. 
One group was led by TP (group-P with KH, AM, NU, and EK) and the second group by FKA (group-K with GH, FA, PDC, FL, GC, and MG). 
We used the same photometric but different spectroscopic data. 
We compared our partial results as the work progressed. 
However, the overall approach used by each group was different. In the end, we combined our results to obtain the final parameters of the system. 
The paper is organized as follows. 
In Sect.\,2 the observational data are introduced. 
The radial velocity and spectral analyses are given in Sect.\,3 and Sect.\,4, respectively. 
The binary modelling and the pulsation frequency analysis are presented in Sect.\,5 and Sect.\,6. 
In Sect.\,7, discussions and conclusions are given.  

\section{Observational Data}

In the photometric analysis of AI\,Hya, \tess\ data was used by both groups. \tess\ was launched in April 2018 mainly to detect new exoplanets \citep{2014SPIE.9143E..20R}. \tess\ has monitored almost the entire sky which has been subdivided into sectors that are observed for about 27 days each. The \tess\ observations were taken in 2-min. short (SC) and 30-min long (LC) cadence in the nominal phase of the mission (first two years). For the extended mission, the LC was reduced to 10-min. The data are available in the Barbara A. Mikulski Archive for Telescopes (MAST)\footnote{https://mast.stsci.edu} where they are released in different versions: simple aperture photometry (SAP) and pre-search data conditioning SAP fluxes (PDCSAP). AI\,Hya was observed in one sector only (sector 7). The 2-min SAP fluxes were used in our analysis since SAP fluxes have lower flux uncertainty and 2-min data are more suitable for the analysis of AI\,Hya (see
Sect.\,\ref{puls_sec}). They were converted into magnitude by using the same method as \cite{2022RAA....22h5003K}.

Photometric data from ground-based surveys also exist, e.g. from ASAS~3 \citep{2002AcA....52..397P} and ASAS-SN \citep{2018MNRAS.477.3145J}, but they are of inferior quality and do not allow for proper analysis of pulsations. The \tess\ sector 7 data are the best ones available so far, although AI\,Hya will again be visible in the satellite's field of view in sector 61.

\begin{table}
\centering
  \caption{Information about the spectroscopic observations. N, R and SNR represent the number of the spectra, resolving power and the signal-to-noise ratio, respectively.}\label{table1}
  \begin{tabular}{@{}lcccccc@{}}
  \toprule
Spectrometer  & N     & Observations    & R         &SNR & Spectral       \\
            &       &  years          &           &    & range [{\AA}]  \\
 \midrule
 CAOS     & 1       & 2021            & 38000     & 50          & 415\,$-$\,670 \\
 CORALIE  & 3       & 2015            & 60000     & 20\,$-$\,34 & 390\,$-$\,680  \\
 HERMES   & 15      & 2020            & 85000     & 50\,$-$\,70 & 377\,$-$\,900  \\
 HIDES    & 13      & 2014\,$-$\,2017 & 50000     & 40\,$-$\,88 & 408\,$-$\,752  \\
\bottomrule
\end{tabular}
\end{table}

The spectroscopic data of the system were taken from four different instruments. The list of the instruments and the basic information about them are given in Table\,\ref{table1}. One spectrum was taken with \textit{Catania Astrophysical Observatory Spectropolarimeter} \citep[CAOS,][]{2016AJ....151..116L}. The CAOS is a high-resolution, fibre-fed, cross-dispersed \'{e}chelle spectrograph installed to the 91-cm telescope at the Catania Astrophysical Observatory (Mt. Etna, Italy). Three spectra of AI\,Hya were collected from the CORALIE  \'{e}chelle spectrograph which is mounted on the 1.2-m Leonhard Euler telescope at La Silla Observatory (Chile) \citep{2018haex.bookE.190P}. 
The \textit{High Efficiency and Resolution Mercator \'{e}chelle spectrograph} (HERMES) was also used to obtain high-resolution spectra of AI\,Hya. 
HERMES is mounted on the 1.2-m Mercator telescope at the Roque de Los Muchchos observatory on the Canary Island La Palma in Spain \citep{2011A&A...526A..69R}.
The last instrument used in this study is the \textit{HIgh-Dispersion \'{E}chelle spectrograph} (HIDES). HIDES is attached to the 1.88-m telescope of Okayama astrophysical observatory in Japan \citep{2013PASJ...65...15K}. 
The spectra of CORALIE and HIDES were taken by group-P, while the spectra of CAOS and HERMES were gathered by group-K. In total 32 spectra of AI\,Hya were gathered and these spectra are well distributed in orbital phases of AI\,Hya. Each group used the obtained spectra to measure the radial velocity (\vr\,) changes. Additionally, these data were taken into account to derive the atmospheric parameters (e.g. effective temperature \teff, surface gravity \logg, metallicity) and the projected rotational velocity (\vsini) of the components of AI\,Hya.

\section{Radial Velocity analysis}
The \vr\, values of the AI\,Hya system were measured with different approaches by both group-P and group-K using different spectra taken from the distinct instruments. 

\subsection{\vr\, measurements}
Group-P calculated the \vr\, values from HIDES and CORALIE spectra, using the two-dimensional cross-correlation 
\textsc{todcor}
program \citep{1994ApJ...420..806Z}. 
In the analysis, a synthetic spectrum was used as a template and this spectrum was generated using
an ATLAS9 model atmosphere \citep{1993KurCD..13.....K} having \teff, metallicity [M/H] and \vsini\, parameters of 6800\,K, 0.0 and 30\,\kms, respectively. When a template with 60\,\kms (the \vsini\, value found in further analysis) was used, the results did not improve in terms of $rms$ of the orbital fit, nor did the uncertainties of orbital elements. Moreover, some points, with the smallest difference in \vr\, measurements, seemed to suffer from systematic effects, and had to be rejected. We therefore believe the use of 30\,\kms templates was justified. The calculated \vr\, values for each binary component are given in Table\,\ref{rvm}.

Group-K used the 
\textsc{RaVeSpAn} 
code \citep{2017ApJ...842..110P} to determine the \vr\, values of the binary components using the broadening function formalism. In the analysis, local thermodynamic equilibrium (LTE) synthetic spectra with atmospheric parameters similar to that of group-P were used as templates \citep{2005A&A...443..735C}. The spectra of CAOS and HERMES were used in the \vr\, measurements. The resulting \vr\, measurements are given in Table\,\ref{rvm}.

\subsection{\vr\, curve modelling}

\begin{figure}
\centering
\includegraphics[width=9cm, angle=0]{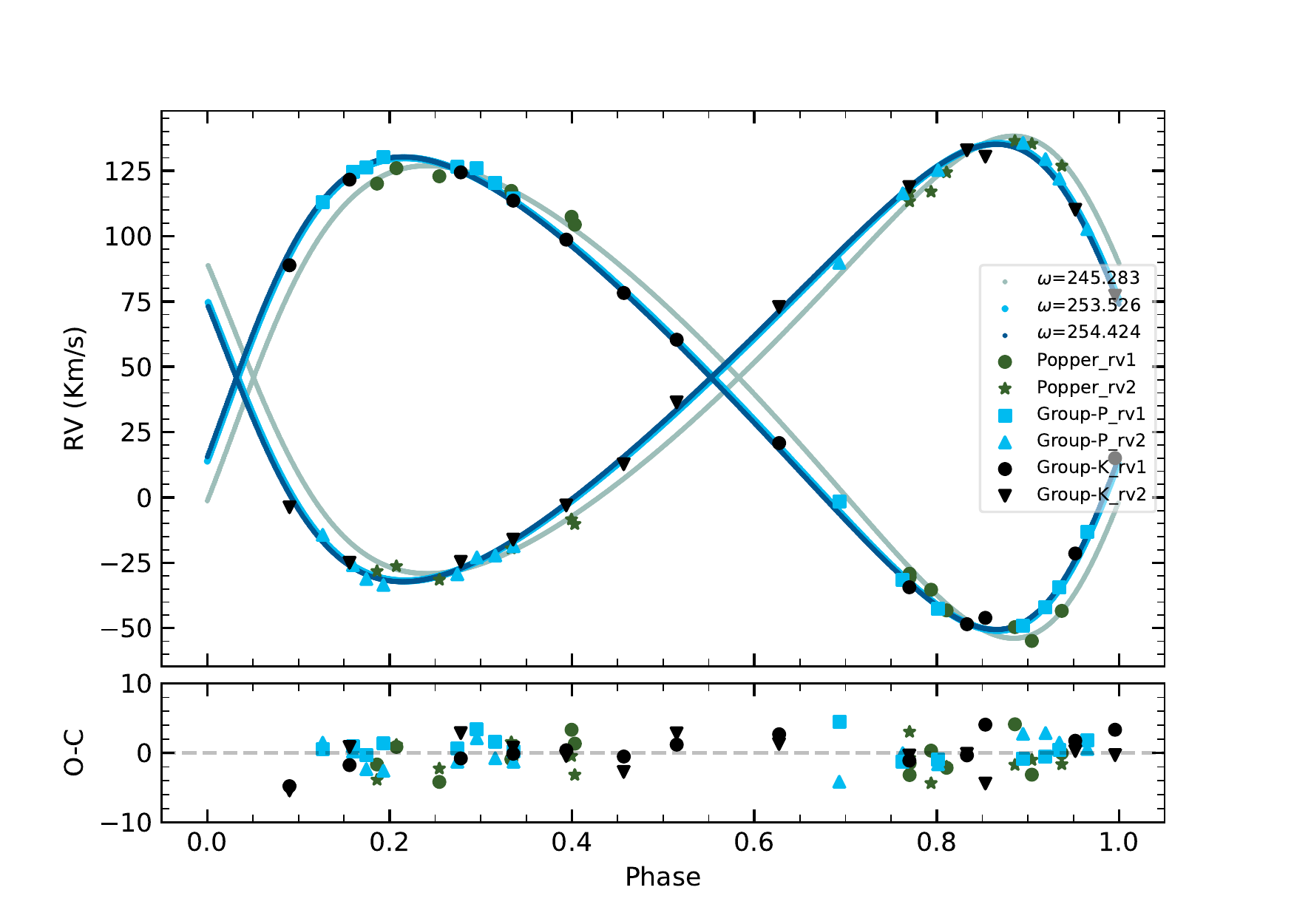}
\caption{Upper panel: The model \vr\, fit to the combined \vr\ measurements from \citet{1988AJ.....95..190P}, Group-P (HIDES+CORALIE) and Group-K (HERMES+CAOS). Lower panel: residuals. Model made by Group-P.}
\label{rv_fit_groupP}
\end{figure}

For the spectroscopic orbital fitting, group-P used all the available \vr\, measurements, including those made by group-K and from \citet{1988AJ.....95..190P}. Group-P used the 
\textsc{v2fit} 
code \citep{2010ApJ...719.1293K} which adjusts a double-Keplerian with a Levenberg-Marquardt algorithm. In this analysis, the amplitude of \vr\, curves ($K$), \porb, the time of phase zero (\tzero), mass centre's velocity ($\gamma$), eccentricity ($e$) and argument of the periastron ($\omega$) were set as free parameters. Thanks to the long time span of the data ($>$51~years), it was possible to detect the apsidal motion ($\dot{\omega}$) of the binary's orbit: 0.186(56)~deg/yr. This is in reasonable agreement (1.75$\sigma$) with the value given by \citet{2020PASJ...72...37L}: 0.075(31)~deg/yr. The results of the analysis are given in Table\,\ref{rvresult} and the theoretical \vr\, curve fits to the measured \vr\, data are illustrated in Fig.\ref{rv_fit_groupP}. 

\begin{figure}
\centering
\includegraphics[width=8.4cm, angle=0]{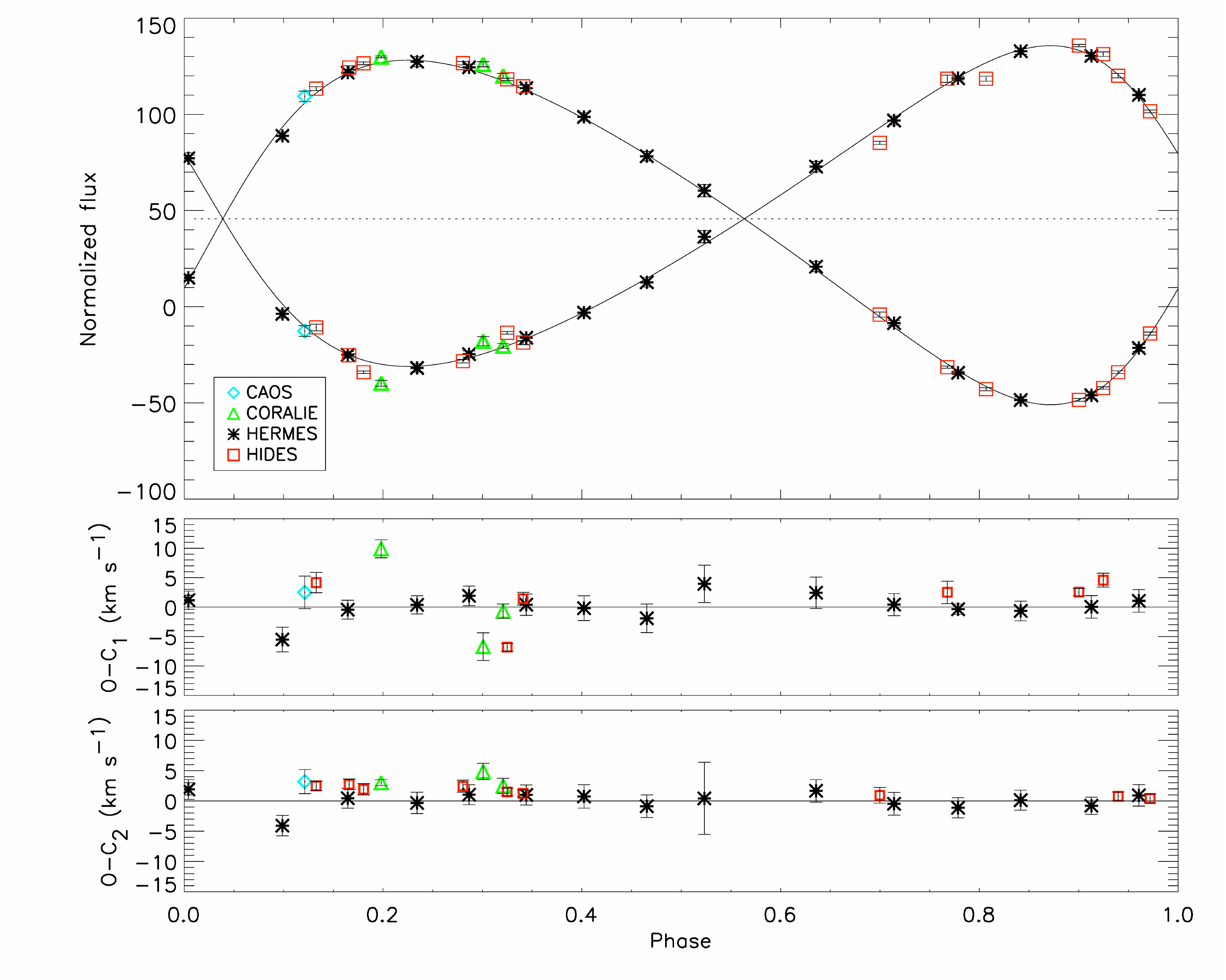}
\caption{Upper panel: The model \vr\, fit to the \vr\, measurements of Groups-K and -P. Lower panel: residuals. Model made by Group-K.}\label{rv_fit_groupK}
\end{figure}

Group-K used the 
\textsc{rvfit} 
code\footnote{http://www.cefca.es/people/riglesias/rvfit html} for the radial velocity analysis. The
\textsc{rvfit} 
program can analyse single and double-lined binary systems by using the adaptive simulated annealing method \citep{2015PASP..127..567I}. 
In the analysis, the \porb\ taken from \cite{2004AcA....54..207K} was considered as a fixed parameter. 
Other orbital parameters such as \tzero, $K$, $\gamma$, $\omega$ and $e$ were taken as free parameters during the analysis. Both groups \vr\, measurements were used in the analysis and as a result, the orbital parameters of the system were obtained. The resulting parameters of the current \vr\, analysis are given in Table\,\ref{rvresult}. The consistency between the theoretical \vr\, curve and measurements is shown in Fig.\,\ref{rv_fit_groupK}.

Both groups found the resulting mass ratio ($q = M_2/M_1 = K_1/K_2$)\footnote{The subscripts 1 and 2 refer to hotter primary and cooler secondary components, respectively.} larger than 1 (1.075\,$\pm$\,0.011 and 1.080\,$\pm$\,0.007 for groups -K and -P, respectively). According to this $q$ value, the \vr\, curve and the results, the star (generally called secondary) covered by the hotter binary component at 
orbital phase 0.5
is more massive than the hotter binary component (primary). To test these findings, binary modelling is necessary. Therefore, these results will be tested in the binary modelling sections.

\begin{table}
\begin{center}
\centering
\caption[]{The results of the radial velocity analysis. The subscripts 1 and 2 refer to hotter primary and cooler secondary components, respectively.
$^a$ shows the fixed parameters.}\label{rvresult}
\begin{tabular}{lrr}
\hline
 Parameter		        & Group-P	                       &  Group-K                  \\
\hline
\tzero\ (HJD)	         & 2458491.570\,$\pm$\,0.028       &2452506.383\,$\pm$\,0.032\\
\porb (d)                   & 8.289761\,$\pm$\,0.000027    &8.2896490$^a$        \\
$\gamma$ (km/s)	         & 45.90\,$\pm$\,0.24	           &45.70\,$\pm$\,0.35         \\
$K_1$ (km/s)		     & 90.42\,$\pm$\,0.37	           &89.52\,$\pm$\,0.65      \\
$K_2$ (km/s)		     & 83.71\,$\pm$\,0.46	           &83.29\,$\pm$\,0.63         \\
$e$			             & 0.2419\,$\pm$\,0.0036           &0.2432\,$\pm$\,0.0050     \\
$\omega$ (deg)		     & 254.03\,$\pm$\,1.30             &250.92\,$\pm$\,1.63      \\
$\dot{\omega}$ (deg/yr)  & 0.186\,$\pm$\,0.056              &   \\
$a_1\sin i$ ($R_\odot$)	 & 14.380\,$\pm$\,0.061            &14.222\,$\pm$\,0.105       \\
$a_2\sin i$ ($R_\odot$)	 & 13.312\,$\pm$\,0.072            &13.233\,$\pm$\,0.101       \\
$a  \sin i$ ($R_\odot$)	 & 27.692\,$\pm$\,0.094            &27.454\,$\pm$\,0.145      \\
$M_1\sin ^3i$ ($M_\odot$)& 1.992\,$\pm$\,0.023             &1.950\,$\pm$\,0.033         \\
$M_2\sin ^3i$ ($M_\odot$)& 2.151\,$\pm$\,0.022             &2.095\,$\pm$\,0.035       \\
$q = M_2/M_1$		     & 1.080\,$\pm$\,0.007             &1.075\,$\pm$\,0.011     \\
 \hline
\end{tabular}
     \end{center}
\end{table}

\section{Spectral analysis}

\subsection{Group-K}
\subsubsection{Spectral disentangling}
To obtain the atmospheric parameters (\teff, \logg), \vsini\, and the chemical composition of each binary component of AI\,Hya, a detailed spectral analysis is necessary. 
As AI\,Hya is a double-lined binary system, its spectrum consists of the spectral lines of both binary components. 
Therefore, group-K carried out a spectral disentangling analysis to extract the individual spectra of each binary component from the composite spectra of AI\,Hya. 
In the analysis, the code \textsc{fdbinary} was used \citep{2004ASPC..318..111I}. 
\textsc{fdbinary} 
is capable of disentangling a composite spectrum, which includes flux contributions from two or three components, in Fourier space. 
Before the analysis with \textsc{fdbinary}, one should know the light contributions of the binary components at the orbital phases corresponding to the times the spectra were taken. These values should be fixed during the analysis. 
Hence, to determine the light contributions of both binary components at the different orbital phases, we carried out a preliminary binary modelling of AI\,Hya by taking \teff\, of the \tess\ Input Catalog \citep[TIC;][]{2019AJ....158..138S} as the \teff\, of the hotter component. The analysis was performed utilizing the 
\textsc{Wilson-Devinney}
code \citep{1971ApJ...166..605W}. As a result of this preliminary analysis, it was found that the hotter and cooler binary components contribute around 38\% and 62\% to the total, respectively. However, one should keep in mind that these light contributions change according to the orbital phases. For example, the primary eclipse is a total eclipse where the light contribution of the hotter components is negligible. 

In the analysis, we used the HERMES spectra as they are well distributed over the orbital phases and have a higher resolving power. Taking into account the observation time of each HERMES 
spectrum, the light contributions at these times were first determined using the fluxes measured from the photometric solution and subsequently fixed during the analysis. 
In addition to this, we also fixed all results derived in the \vr\, analysis during the spectral disentangling. For the disentangling progress, we used the spectral interval of $\sim$4200\,$-$\,6400\,{\AA} by ignoring the parts polluted by telluric lines. For the analysis, this spectral window was divided into 15 spectral parts with 
steps of $\sim100-150$\,{\AA}. 
Each small spectral part 
was
then analysed separately. As a result, we obtained the individual spectra of each binary component. 
The separated spectra derived with \textsc{fdbinary} were re-normalised by taking into account the light ratio of the binary components, as described by \cite{2004ASPC..318..111I}. 

\subsubsection{Determination of the atmospheric parameters and chemical compositions}

After the individual spectra of the components of AI\,Hya were obtained, we
were able to determine the atmospheric parameters, \vsini, and the chemical composition.
To derive these parameters, we used the plane-parallel and line-blanketed local thermodynamic equilibrium (LTE) ATLAS9 model atmospheres \citep{1993KurCD..13.....K} and the 
\textsc{synthe} 
code \citep{1981SAOSR.391.....K} to generate theoretical spectra. First, the hydrogen lines of the binary components were used to obtain initial \teff\ values.

\begin{figure}
\centering
\includegraphics[width=8cm, angle=0]{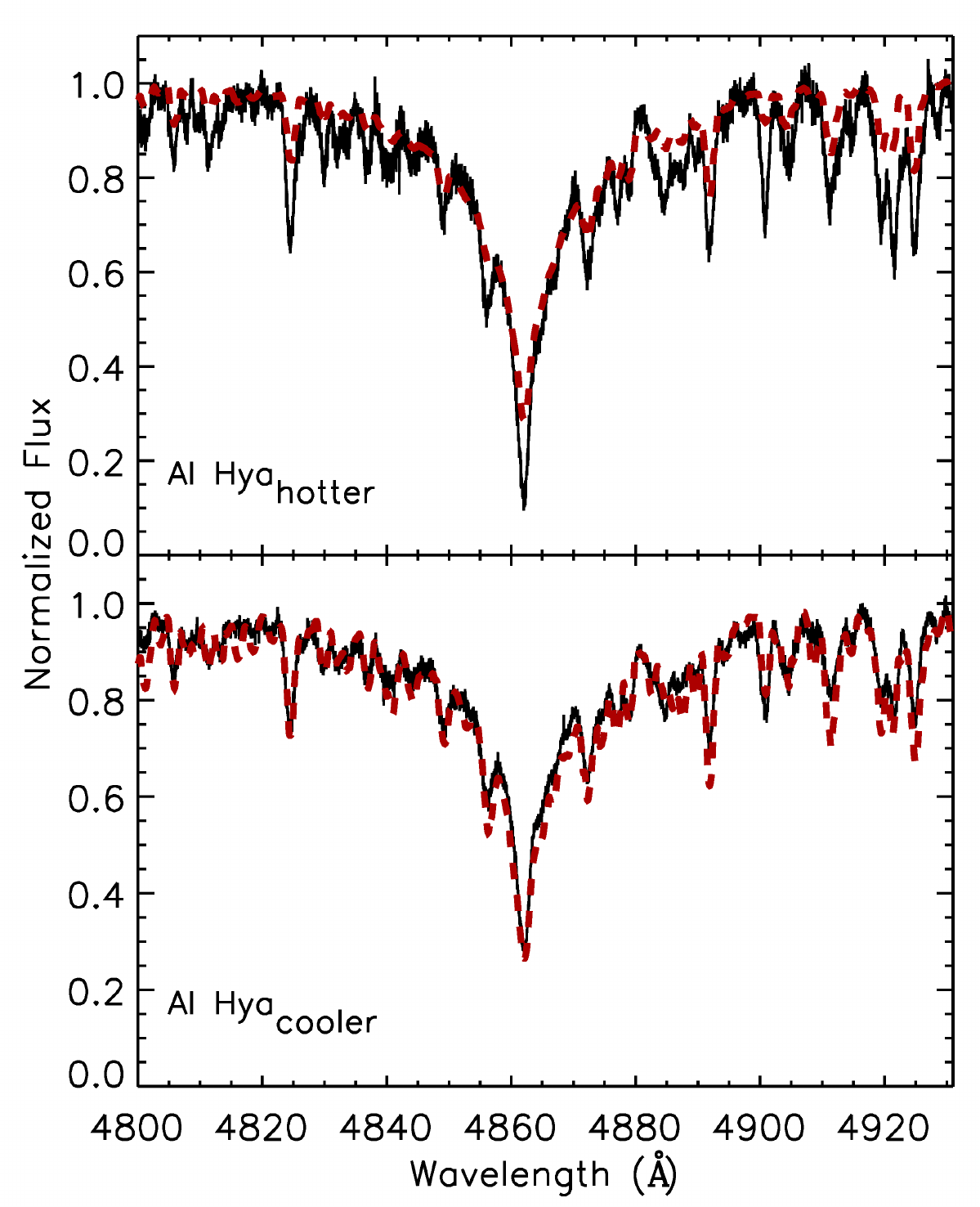}
\caption{Theoretical hydrogen line fits (red dashed lines) to the $H_\beta$ lines (solid black line) of the hotter and cooler binary components (Group-K).}
\label{hlinefit}
\end{figure}

In this analysis, the $H_\beta$ lines of the components were compared with many theoretical $H_\beta$ lines which were derived for a wide range of \teff\, ($5000-9000$\,K) with a step size of 100\,K, where \logg\, and metallicity were fixed to 4.0 and solar, respectively. 
During the analysis, we took into account the minimization method described by \cite{2004A&A...425..641C} and successfully applied in a series of papers \citep[i.e.,][]{2022MNRAS.515.4350C, 2019MNRAS.484.2530C}. Consequently, the \teff\, of the hotter and cooler components were found to be 7500\,$\pm$\,200\,K and 7000\,$\pm$\,150\,K, respectively. We did not attempt to optimize \logg\ because the hydrogen lines are not sensitive to this parameter for stars cooler than 8000\,K \citep{2002A&A...395..601S}. 
The best theoretical $H_\beta$ line fits to the separated spectra of the components are shown in Fig.\,\ref{hlinefit}.

We also determined 
values for \logg, the microturbulent velocity $\xi$, and \vsini\ 
by improving the initially determined \teff\, value using the excitation potential$-$abundance relationship. For the correct atmospheric parameters, different excitation potentials of the same element should give the same abundances. Therefore, by using this relation for iron (Fe), we determined the atmospheric parameters. Detailed information about this analysis
method is given by \cite{2016MNRAS.458.2307K}. The results of this analysis are listed in Table\,\ref{atmpar_result}. To determine the errors 
on 
the atmospheric parameters, we checked how their values change for differences in the excitation potential$-$abundance correlation of about 5\%.

\begin{figure}
\centering
\includegraphics[width=8cm, angle=0]{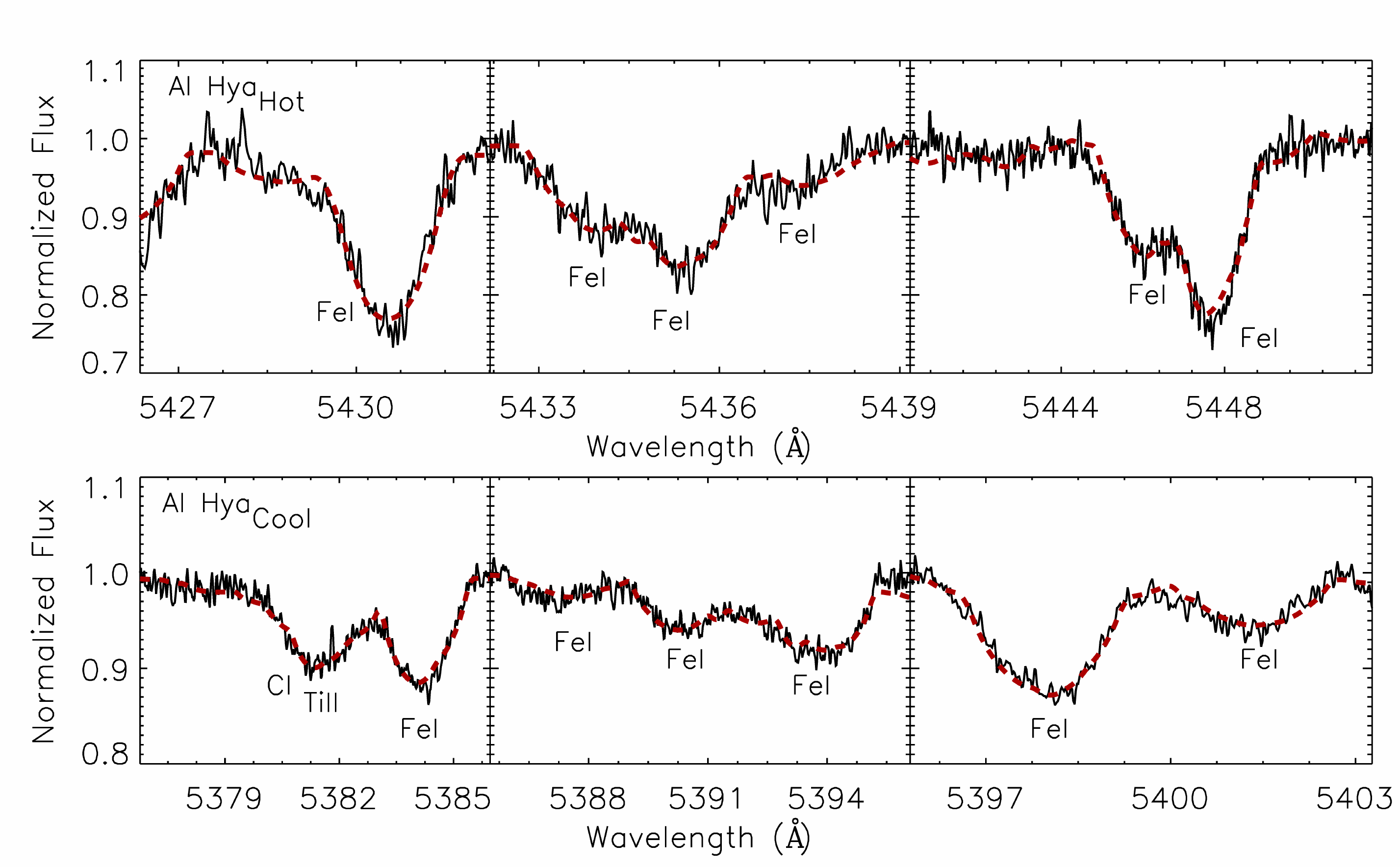}
\caption{Consistency between the synthetic (dashed-lines) and disentangled spectra of the components of AI\,Hya (Group-K).}\label{abunfit}
\end{figure}

\begin{figure}
\centering
\includegraphics[width=8cm, angle=0]{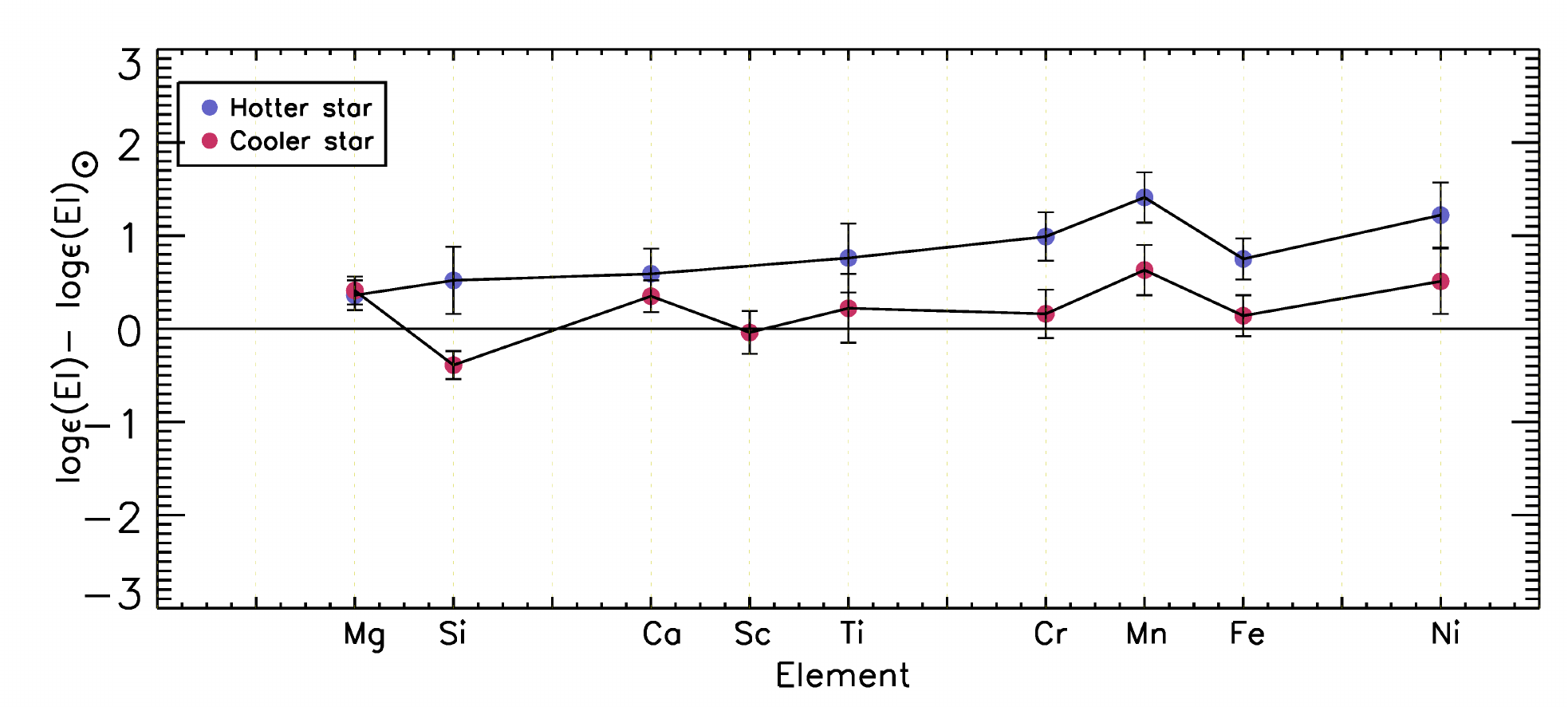}
\caption{Abundance distribution of the components of AI\,Hya relative to solar values \citep{2009ARA&A..47..481A} (Group-K).}
\label{abundist}
\end{figure}

In the next step, the chemical composition of the binary components was derived after fixing the atmospheric parameters to their final values. For the chemical abundance determination, we first identified the lines based on the Kurucz line list\footnote{http://kurucz.harvard.edu/linelists.html}. 
The spectral synthesizing method and the identified lines were used in this examination. Consequently, the chemical compositions of both binary components were obtained and the results are listed in Table\,\ref{abunresult}. The consistency between the synthetic and observed spectra of both binary components is illustrated in Fig.\,\ref{abunfit}. The abundance distributions relative to solar abundance \citep{2009ARA&A..47..481A} are shown in Fig.\,\ref{abundist}, indicating that the hotter binary component has an overabundance compared to the Sun for some elements. The errors of the chemical compositions were determined including the uncertainties in the derived atmospheric parameters and the effects of the resolving power and the SNR of the spectra, as described by \cite{2016MNRAS.458.2307K}.

\begin{table*}
\centering
\caption{The final atmospheric parameters and $v \sin i$ value of the hot (primary) and cool binary components of AI\,Hya. $\log \epsilon$ (Fe) represent the relative abundance with respect to hydrogen (H=12.0)}. \label{atmpar_result}
\begin{tabular}{llcccc}
\toprule
                    \multicolumn{6}{c}{\hrulefill \,Group-K\,\hrulefill}\\
&$T_{\rm eff}$\,(K)     & $\log g$\,(cgs)     & $\xi$\,(km\,s$^{-1}$)   & $v \sin i$\,(km\,s$^{-1}$) & $\log \epsilon$ (Fe)\\
\hline

Primary  &7700\,$\pm$\,100 & 3.8\,$\pm$\,0.1 & 3.4\,$\pm$\,0.3 & 57\,$\pm$\,6 & 8.25\,$\pm$\,0.54\\
Secondary &7200\,$\pm$\,100 & 3.6\,$\pm$\,0.2 & 1.9\,$\pm$\,0.3 & 64\,$\pm$ 4 & 7.64\,$\pm$\,0.20\\
                     \hline
                     \multicolumn{6}{c}{\hrulefill \,Group-P (\textsc{gssp})\,\hrulefill}\\
                     &$T_{\rm eff}$\,(K)     & $\log g$\,(cgs)     & $\xi$\,(km\,s$^{-1}$)   & $v \sin i$\,(km\,s$^{-1}$) & $\mathrm{[M/H]}$  \\
\hline
Primary  &7350\,$\pm$\,300 & 3.8 (fixed) & 4.83\,$\pm$\,1.15 & 50 (fixed) & 0.14\,$\pm$\,0.14\\
Secondary &7150\,$\pm$\,250 & 3.6 (fixed) & 3.07\,$\pm$\,0.52 & 62 (fixed) & 0.06\,$\pm$\,0.10\\
                     \hline
                     \multicolumn{6}{c}{\hrulefill \,Group-P (\textsc{iSpec})\,\hrulefill}\\
                     &$T_{\rm eff}$\,(K)     & $\log g$\,(cgs)     & $\xi$\,(km\,s$^{-1}$)   & $v \sin i$\,(km\,s$^{-1}$) & $\mathrm{[M/H]}$  \\
                     \hline
                     Primary  &7300\,$\pm$\,170 & 3.83 (fixed) & 5.33\,$\pm$\,0.86 & 50 (fixed) & 0.15 (fixed)\\
                     Secondary &7260\,$\pm$\,175 & 3.58 (fixed) & 3.98\,$\pm$\,0.70 & 62 (fixed) & 0.01 (fixed)\\
                      \hline
\bottomrule
\end{tabular}
\end{table*}

\begin{table}
\begin{center}
\caption[]{Abundances of individual elements of the binary components and Sun (\citealt{2009ARA&A..47..481A}).}\label{abunresult}
 \begin{tabular}{lccc}
 \toprule
 \multicolumn{4}{c}{\hrulefill \,Group-K\,\hrulefill}\\
 \hline
  Elements & Hotter  & Cooler      &Solar             \\
           &component& component   &abundance\\
  \hline\noalign{\smallskip}
$_{12}$Mg &7.96\,$\pm$\,0.16     & 8.01\,$\pm$\,0.63      & 7.60\,$\pm$\,0.04\\
$_{14}$Si &8.03\,$\pm$\,0.36     & 7.12\,$\pm$\,0.51      & 7.51\,$\pm$\,0.03\\
$_{20}$Ca &6.93\,$\pm$\,0.27     & 6.69\,$\pm$\,0.27      & 6.34\,$\pm$\,0.04\\
$_{21}$Sc &                      & 3.11\,$\pm$\,0.32      & 3.15\,$\pm$\,0.04\\
$_{22}$Ti &5.71\,$\pm$\,0.49     & 5.17\,$\pm$\,0.30      & 4.95\,$\pm$\,0.05\\
$_{24}$Cr &6.63\,$\pm$\,0.42     & 5.80\,$\pm$\,0.30      & 5.64\,$\pm$\,0.04\\
$_{25}$Mn &6.84\,$\pm$\,0.82     & 6.06\,$\pm$\,0.45      & 5.43\,$\pm$\,0.05			   \\
$_{26}$Fe &8.25\,$\pm$\,0.23     & 7.64\,$\pm$\,0.24      & 7.50\,$\pm$\,0.04\\
$_{28}$Ni &7.44\,$\pm$\,0.38     & 6.73\,$\pm$\,0.33      & 6.22\,$\pm$\,0.04\\
  \noalign{\smallskip}\hline
      \multicolumn{4}{c}{\hrulefill \,Group-P (\textsc{iSpec})\,\hrulefill}\\
\hline
        Elements & Hotter  & Cooler      &Solar             \\
           &component& component   &abundance\\
\hline
$_{24}$Cr &5.95\,$\pm$\,0.19    & 5.63\,$\pm$\,0.23   & 5.64\,$\pm$\,0.04\\
$_{26}$Fe &7.83\,$\pm$\,0.16    & 7.48\,$\pm$\,0.17   & 7.50\,$\pm$\,0.04\\
$_{28}$Ni &6.76\,$\pm$\,0.18    & 6.53\,$\pm$\,0.22   & 6.22\,$\pm$\,0.04\\
\bottomrule
\end{tabular}
\end{center}  
\end{table}

\subsection{Group-P}

For the spectral decomposition and analysis, group-P used the HIDES data only. Spectral analysis was performed on both the observed composite spectra and the disentangled spectra of the individual components. 
For the spectral disentangling, we used a python wrapper\footnote{https://github.com/ayushmoharana/fd3\_initiator} made for using version 3 of 
\textsc{fdbinary} \citep[FD3;][]{2004ASPC..318..111I}. A particular portion of the total spectra was taken to ensure good quality in terms of SNR and spectral features. The light fractions used for the disentangling procedure were obtained from the light curve analysis as 38\% and 62\% for the primary and secondary respectively. 

\subsubsection{\textsc{gssp}}
On the other hand, we also modelled the composite spectrum using the {\footnotesize \textsc{gssp\_composite}} module of the Grid Search in Stellar Parameter (\textsc{gssp}) software package \citep{2015A&A...581A.129T}. As its name implies, \textsc{gssp} is based on a grid search in the fundamental atmospheric parameters. It uses the method of atmosphere models and spectrum synthesis, which performs a comparison of the observations with theoretical spectra from the grid. These synthetic spectra are calculated using the \textsc{synthV} LTE-based radiative transfer code \citep{1996ASPC..108..198T} and a grid of atmospheric models pre-computed using \textsc{llmodels} \citep{2004A&A...428..993S}. Specifically, in the composite module, the user can set the radial velocity of the components as a free parameter so that all the possible combinations of the synthetic spectra of primary and secondary from the computed grid are used to build the composite theoretical spectra of the binary. This synthetic spectrum is then compared against the a-priori normalized observed spectrum and a $\chi^2$ merit function is used to judge the goodness of the fit.

\begin{figure}
    \centering
    \includegraphics[width=8cm]{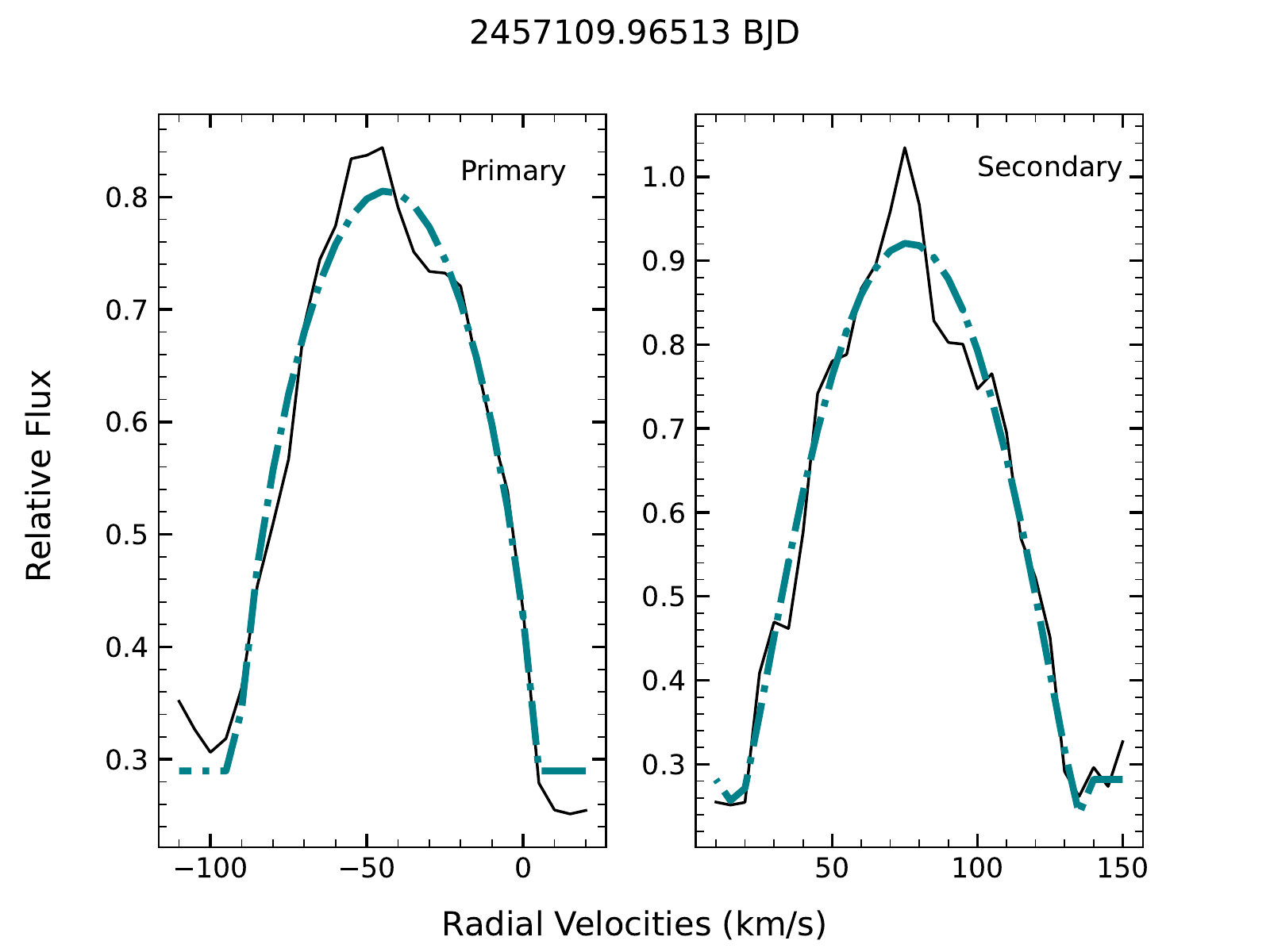}
    \caption{Broadening functions for the primary and secondary components of AI Hydrae calculated using HIDES spectra (epoch: 2457109.96513\,HJD), which provided a good SNR and velocity separation between the two components. The blue, dashed line represents best-fit rotational function (Group-P).}
    \label{fig:BF_AIHya}
\end{figure}

The broadening function (BF) is a representation of spectral profiles in velocity space. The BF contains signatures of the 
\vr\
shifts of different lines and also intrinsic stellar effects like rotational broadening, spots, pulsations, etc. \citep{1999TJPh...23..271R}. We calculated the BF for one of the composite spectra of AI Hya to estimate $v \sin i$ values for the primary and secondary components, respectively. This process serves to remove the degeneracy between $v \sin i$ and other atmospheric parameters like $T\mathrm{_{eff}}$ and [M/H]. A modified version of the treatment described in \citet{1999ASPC..185...82R} was adopted and a multi Gaussian fit was implemented. The BF was calculated in a wavelength range of 4080-5000 \AA. A synthetic solar-type spectrum with zero projected rotational velocity $v \sin i$ was used as our template. To deal with the noise in the data, a Gaussian smoother of 3\,$\mathrm{km\,s^{-1}}$ rolling window was applied to the BF. Two clear peaks were visible in the velocity space, as shown in Figure \ref{fig:BF_AIHya}, corresponding to the primary and secondary components. The peaks were fitted with the rotational profile,
\begin{equation}
    G(v)=A \left [c_{1}\sqrt{1-\left(\frac{v}{v_{max}}\right)^{2}}+c_{2}\left (1-\left (\frac{v}{v_{max}}\right)^{2}\right )\right] +l v+k 
\end{equation}
where $A$ is the area under the profile, $v_{max}$ is the maximum velocity shift which occurs at the equator \citep{2005oasp.book.....G}, $c_{1}$ and $c_{2}$ are constants which are a function of limb darkening themselves, while $l$ and $k$ are correction factors to the BF continuum. The BF fit was calculated for the spectra with the highest SNR and good separation between the components in velocity space. The best BF fit to the line profile of the primary and secondary binary components are shown in Fig.\ref{fig:BF_AIHya}. Fixing the obtained values of $v \sin i$ from this analysis and $\log g$ from the light curve solution, the \textsc{gssp\_composite} fitting routine was applied to obtain stellar temperatures \teff$\mathrm{_{(1,2)}}$, microturbulent velocities $\xi$ and global metallicities [M/H].

\begin{figure}
\centering
\includegraphics[width=8cm, angle=0]{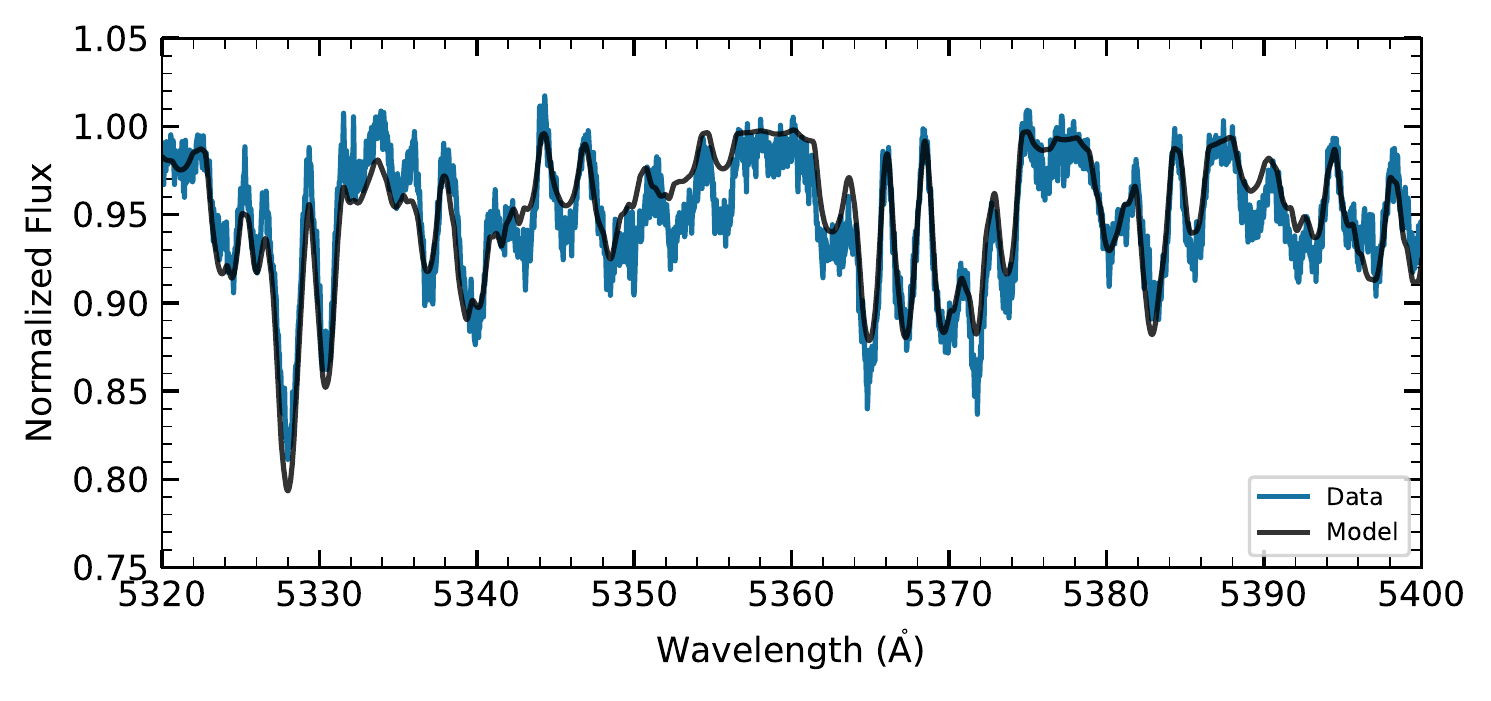}
\caption{A snippet of the best-fit model generated by \textsc{gssp} for the given set of parameters (Group-P).}
\label{fig:gsspfit}
\end{figure}

The step size of the grid gives us a rough idea of the errors involved. However, to obtain more robust error estimates we plotted the $\chi^2$ data for each parameter and fitted a parabola to obtain the minimum; its distance to the intercepts on the abscissa are taken as the errors. These parameters are obtained for a total of four spectra and then averaged out. The remaining spectra were not suitable for the analysis in \textsc{gssp} due to lower SNR. The results of the analysis are compiled in Table \ref{atmpar_result} and a sample of the fit to one of the spectra is shown in Figure \ref{fig:gsspfit}.

\subsubsection{\textsc{iSpec}}
A complimentary spectroscopic analysis was performed on the disentangled spectra of the primary and secondary stars using \textsc{iSpec} \citep{2014A&A...569A.111B}. Before the analysis, the spectra are
treated for 
\vr\
offset and continuum correction. Estimates of flux errors were introduced as a sum of errors calculated from SNR, and flux-scaled residuals from the disentangled routine. For the spectroscopic analysis we fixed the \logg\, parameter with values obtained from the light curve solution and limb darkening parameters with values adopted from \citet{2011A&A...529A..75C}.

We fit the model using the spectral synthesis approach. This is done by implementing the use of 
the \textsc{spectrum} code \citep{1994AJ....107..742G}, a
\textsc{marcs} \citep{2008A&A...486..951G} grid of model atmospheres, and solar abundances taken from \citet{2009ARA&A..47..481A}. We adopt a two-step process. The initial run is aimed at estimating the global metallicity ([M/H]) by keeping it as a free parameter. The macroturbulent
velocity (v$_{mac}$) and alpha enhancement parameters were set to zero as v$_{mac}$ has a negligible contribution for stars in the concerned temperature range and alpha enhancement, when set as a free parameter, produced implausible values. $v \sin i$ was set to 
the values obtained by the BF analysis. We compared the obtained value for [M/H] with results from 
the \textsc{gssp} analysis and found it to be consistent with the
errors. The average value of [M/H] was calculated and fixed for the next step where we fit for temperature \teff, microturbulent velocity $\xi$, and abundances of Iron (Fe), Nickel (Ni) and Chromium (Cr), as these were the prominent lines in the chosen spectral range.

The output parameters obtained from \textsc{iSpec} are given in Table\,\ref{atmpar_result} and Table\,\ref{abunresult}. It is to be noted that Fe, Ni, and Cr are more abundant in the primary compared to solar values
and those of the secondary star.
This trend in the abundances is in agreement with the values obtained by group-K. The output parameters for the secondary star agree fairly well with those
from the \textsc{gssp} analysis and from the group-K. 
The best fit solution for the primary component, as in the case of \textsc{gssp} analysis, also hinted towards a lower $T_\mathrm{eff}$ compared to the group-K solution.



\section{Binary modelling}
\subsection{Group-K}

To update the fundamental stellar parameters ($M$, $R$) of AI\,Hya, we performed binary modelling with the help of the determined atmospheric parameters and the results of the \vr\, investigation.

In binary modelling, the \tess\ data were used. However, 
the shapes of the eclipses of AI\,Hya are distorted due to the pulsations. 
Thus we first cleaned the pulsations and only then carried out the binary modelling. Therefore, the \textsc{Period04} program \citep{2005CoAst.146...53L} was used to detect the variations caused by oscillations. The derived pulsation frequencies\footnote{The frequencies given in Sect.\,6.} were cleaned from the light curve and the residuals were used in the binary modelling. 

In this analysis, we used the 
\textsc{Wilson-Devinney} 
code \citep{1971ApJ...166..605W} combined with Monte-Carlo simulations \citep{2004AcA....54..299Z, 2010MNRAS.408..464Z}. The pulsation removed data were binned to around 4000 points to be used in the binary modelling code. AI\,Hya is classified as a detached binary system in the literature \citep{2020PASJ...72...37L}. According to their results (e.g., for $\Omega$, $q$, $a$), both components do not seem to fill their Roche lobe, hence the system is defined as a detached binary. 
Also, the morphology of the light curve, i.e. very small ellipsoidal variations and eclipses spanning a small fraction of the orbital period, confirm this classification.
 Therefore, a detached binary configuration was considered our analysis. In the modelling, we took some parameters fixed, such as the \teff\, of the hotter component, 
\porb, 
$q$ taken from our results and bolometric albedos \citep{1969AcA....19..245R}, bolometric gravity-darkening coefficient \citep{1924MNRAS..84..665V}, and the logarithmic limb darkening coefficient \citep{1993AJ....106.2096V} taken the same as given \cite{2019MNRAS.488.5279K}. The orbital inclination ($i$), \teff\, of the cooler component, phase shift ($\phi$), $e$, $a$, $\omega$, 
and dimensionless potential ($\Omega$) of the components were set free.

\begin{table}
\begin{center}
\caption[]{Results of the light curve analysis and the fundamental stellar parameters. The Subscripts 1, 2 and 3 represent the hotter, the cooler, and third binary components, respectively. $^a$ Shows the Fixed Parameters.}\label{lcresult}
 \begin{tabular}{lrr}
  \hline\noalign{\smallskip}
   Parameter  &  Value     & Value \\
  \hline\noalign{\smallskip}
                                   & Group-K               & Group-P \\
\hline
$i$ ($^{o}$)                       & 89.866\,$\pm$\,0.015  &   89.837\,$\pm$\,0.136\\
$T$$_{1}$$^a$ (K)                  & 7700\,$\pm$\,100      & 7330\,$\pm$\,170\\
$T$$_{2}$ (K)                      & 7180\,$\pm$\,230      & 7210\,$\pm$\,150\\
$\Omega$$_{1}$                     & 11.412\,$\pm$\,0.046  & -\\
$\Omega$$_{2}$                     & 8.961\,$\pm$\,0.035   &  -\\
Phase shift                        & -0.0310\,$\pm$\,0.0001&  -\\
$q$                                & 1.074$^a$             & 1.075\\
$r$$_{1}$$^*$ (mean)               & 0.1001\,$\pm$\,0.0036 & 0.1015\,$\pm$\,0.0005\\
$r$$_{2}$$^*$ (mean)               & 0.1412\,$\pm$\,0.0026 & 0.1412\,$\pm$\,0.0006\\
$l$$_{1}$ / ($l$$_{1}$+$l$$_{2}$)  & 0.381\,$\pm$\,0.016   & 0.374\,$\pm$0.02\\
$l$$_{2}$ / ($l$$_{1}$+$l$$_{2}$)  & 0.619\,$\pm$\,0.016   & 0.616\,$\pm$\,0.02\\
$l$$_{3}$                          & 0.0                   & 0.0 \\
\multicolumn{2}{c}{Derived Quantities}\\
$M$$_{1}$ ($M_\odot$)              & 1.950\,$\pm$\,0.033  & 1.950\,$\pm$\,0.033\\
$M$$_{2}$ ($M_\odot$)              & 2.096\,$\pm$\,0.035  & 2.096\,$\pm$\,0.035\\
$R$$_{1}$ ($R_\odot$)              & 2.754\,$\pm$\,0.015  & 2.787\,$\pm$\,0.020\\
$R$$_{2}$ ($R_\odot$)              & 3.863\,$\pm$\,0.021  & 3.877\,$\pm$\,0.026\\
log ($L$$_{1}$/$L_\odot$)          & 1.381\,$\pm$\,0.034  &     1.311\,$\pm$\,0.081 \\
log ($L$$_{2}$/$L_\odot$)          & 1.554\,$\pm$\,0.035  &     1.549\,$\pm$\,0.097 \\
$\log g$$_{1}$ (cgs)               & 3.848\,$\pm$\,0.003  &     3.838\,$\pm$\,0.005 \\
$\log g$$_{2}$ (cgs)               & 3.586\,$\pm$\,0.003  &     3.582\,$\pm$\,0.005\\
$M_{bol}$$_{1}$ (mag)              & 1.30\,$\pm$\,0.08    &    1.474\,$\pm$\,0.202\\
$M_{bol}$$_{2}$ (mag)              & 0.87\,$\pm$\,0.08    &    0.877\,$\pm$\,0.243\\
$M_{V}$$_{1}$ (mag)                & 1.25\,$\pm$\,0.08    &    1.424\,$\pm$\,0.208\\
$M_{V}$$_{2}$ (mag)                & 0.79\,$\pm$\,0.08    &   0.822\,$\pm$\,0.258\\
 Distance (pc)                     & 659  $\pm$ 30        &  642 $\pm$ 36\\          
  \noalign{\smallskip}\hline
\end{tabular}
\end{center}
\end{table}

\begin{figure}
\centering
\includegraphics[width=8cm, angle=0]{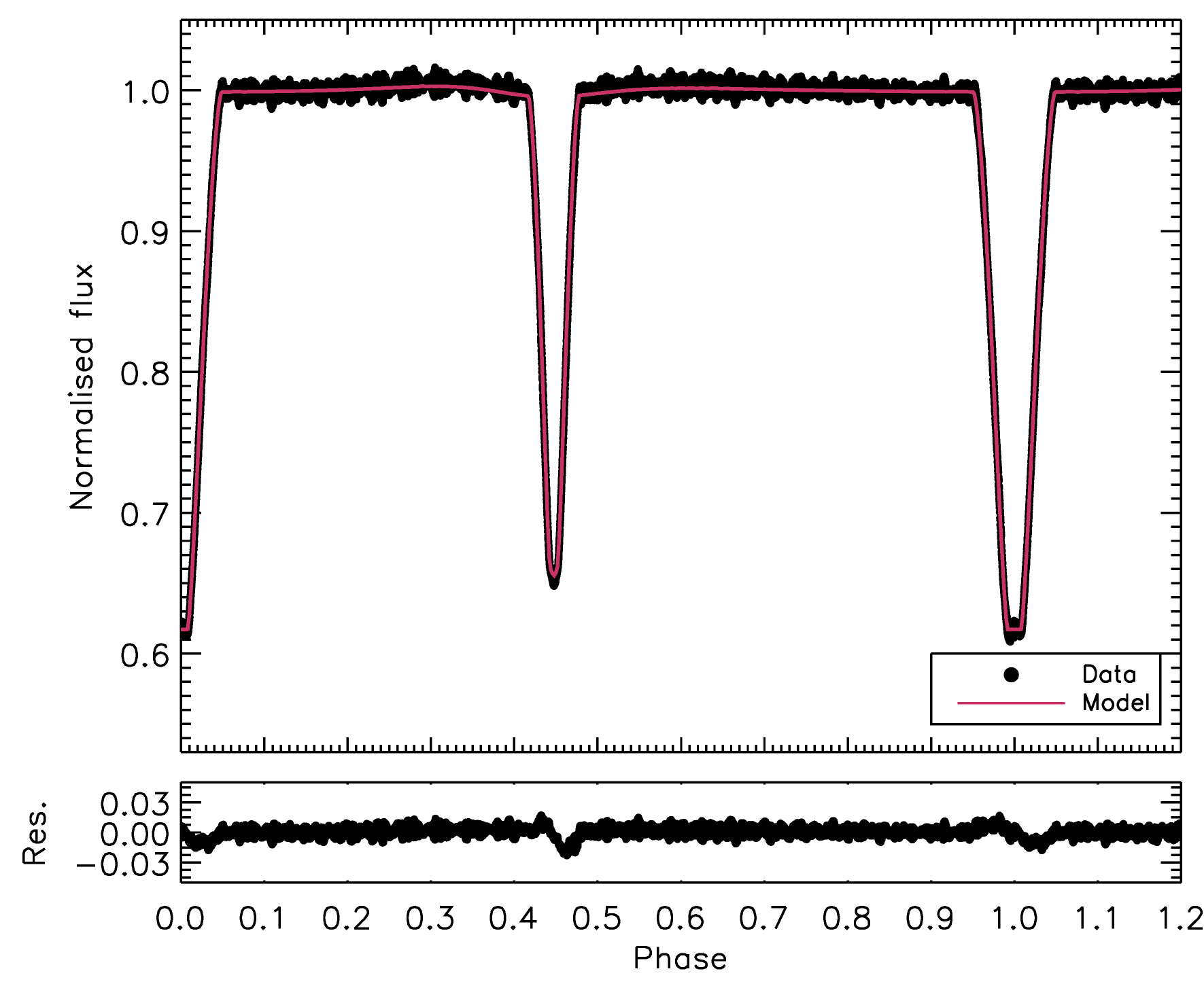}
\caption{Theoretical binary modelling fit without spot assumption (solid-line) (Group-K).}
\label{lccurve}
\end{figure}

As a result of this analysis, the fundamental parameters of both components of AI\,Hya were calculated. Additionally, the bolometric ($M_{bol}$) and absolute ($M_{V}$) magnitudes were estimated. The \textsc{jktabsdim}
code \citep{2004MNRAS.351.1277S} and the bolometric correction \citep{2020MNRAS.496.3887E} are used in the calculations of these parameters. The outcome of the binary modelling is given in Table\,\ref{lcresult} and the consistency of the theoretical light curve with the observation is shown in Fig.\,\ref{lccurve}.

When the results of this analysis were examined, one can notice that the more luminous star is the more massive and also the cooler component. This result is consistent with the results found in the \vr\, analysis by group-K.


\subsection{Group-P}

Aiming to determine precise physical and orbital parameters of AI Hya, we performed its modelling in version 40 of the \textsc{jktebop} \citep{2004MNRAS.351.1277S}. 
This program is written by J. Southworth and aimed at modelling light curves of detached eclipsing binaries and is based on the 
\textsc{ebop} 
program \citep{1981AJ.....86..102P}. The code treats stars as spheres to calculate the eclipse shapes, and biaxial ellipsoids to calculate proximity effects. The light curves are calculated by numerical integration of concentric circles over each stellar surface. It can deal with stellar oblateness of up to 4\% making it a good choice for AI Hya. The photometric data remain the same as used by Group-K. 

The parameters set as free are 
\porb,
time of minima of the primary eclipse $T_o$, inclination $i$, eccentricity $e$, argument of periastron $\omega$, surface brightness ratio $J$ (secondary/primary), ratio of radii ($\frac{\mathrm{r{_A}}}{\mathrm{r{_B}}}$), and the sum of radii (r$\mathrm{_A}$+r$\mathrm{_B}$). These radii are relative to the semi-major axis. For the limb darkening coefficients, we use a logarithmic law and set their initial values according to \citet{2017A&A...600A..30C}. The coefficients were fixed for the initial fit and were perturbed at the error estimation step.

\begin{figure}
\centering
\includegraphics[width=9cm, angle=0]{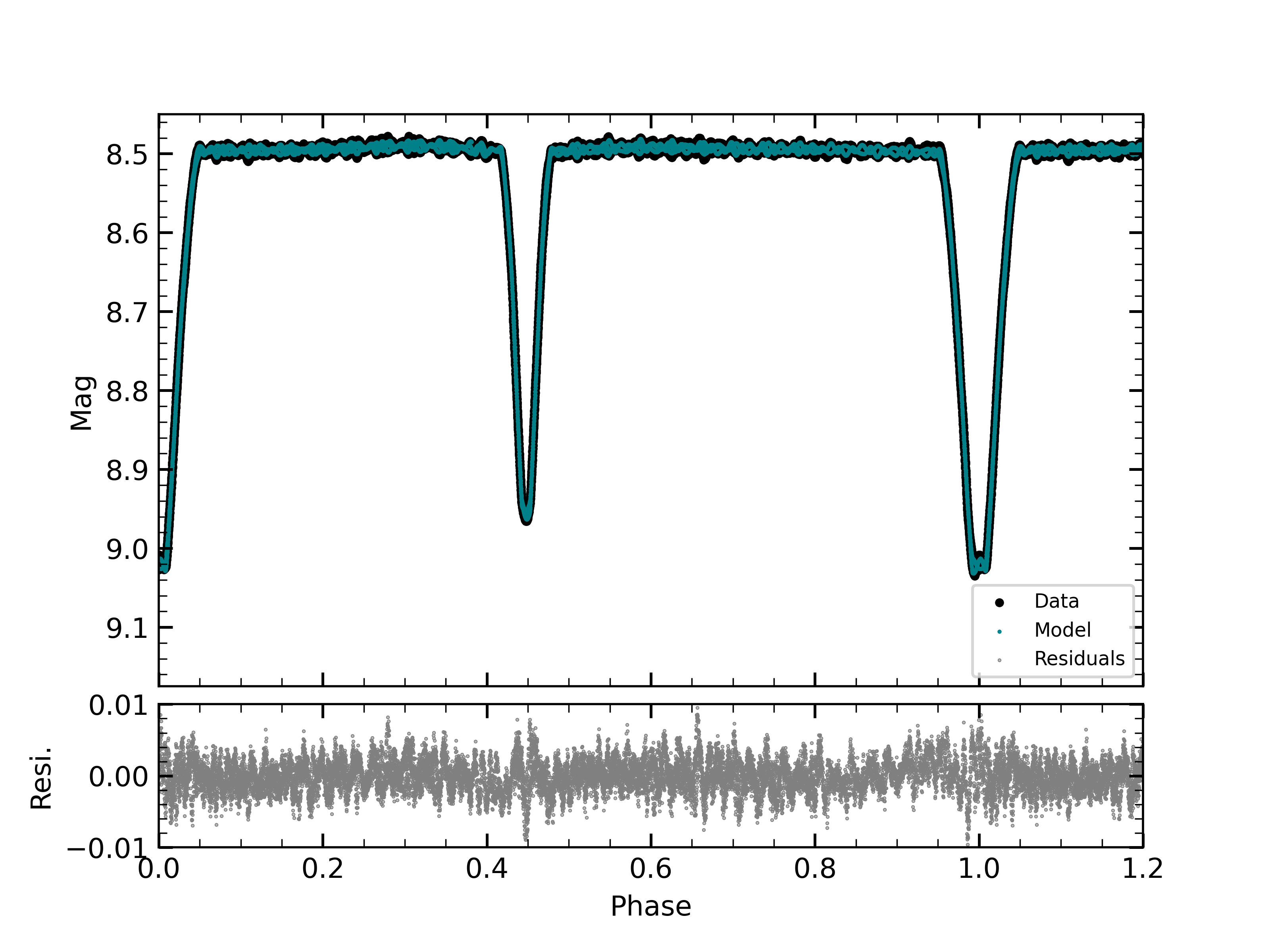}
\caption{
\textsc{jktebop} 
model with 9 sines used to model the pulsations (Group-P).}
\label{fig:lcmodelwsine}
\end{figure}

\begin{figure}
\centering
\includegraphics[width=8.5cm, angle=0]{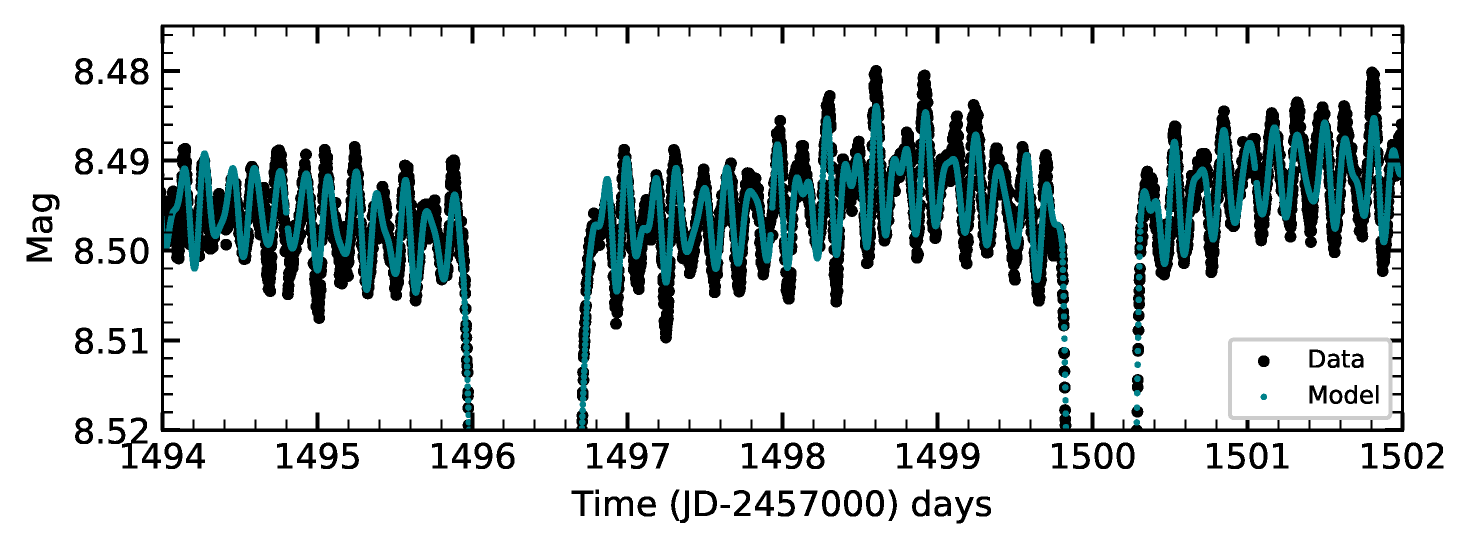}
\caption{Zoomed-in view of the model over an orbit (Group-P).}
\label{fig:lcmodelpuls}
\end{figure}

The code gives an option to include multiple sine and polynomial functions during the light curve modelling to account for periodic and long-term trends. We use this functionality to our advantage to pseudo-model the observed pulsations so that their effect on the binary model is minimal, giving us an improved precision. We analyse the out-of-eclipse portions of the light curve using \textsc{pyriod}\footnote{https://github.com/keatonb/Pyriod}, and use the frequencies to initialise the sinusoids in the 
\textsc{jktebop} 
input files. This is done in an iterative way where we add one sine with a constant period and fit for its epoch and amplitude. The frequency is kept if the model is improved significantly; otherwise the next most prominent frequency is taken. In this analysis, we used a total of 9 sines, which is the limit for 
\textsc{jktebop}. 
The number of independent frequencies of AI Hya is higher than this maximum limit, hence we are left with some residual pulsation signals as seen in Figure \ref{fig:lcmodelwsine} and Figure \ref{fig:lcmodelpuls}.

Once the sines are fixed to the best fit values of epoch, period and amplitudes, we make the Monte Carlo runs for error estimation. The results of this analysis are mentioned in Table \ref{lcresult}, in comparison to the values obtained by group-K. Similarly to the other group, we used the results of \vr, and 
\textsc{jktebop} 
solutions to calculate a set of absolute parameters, including masses, radii, luminosities, and distance. The effective temperatures mentioned in the table are an average over the sum of $T_\mathrm{eff}$ obtained from \textsc{gssp} and \textsc{iSpec} analysis.


\section{Frequency analysis of the pulsations} \label{puls_sec}


AI\,Hya was observed by \tess\ during observation sector 7 in January/February 2019. We used the Simple Aperture Photometry data from the 2-min cadence light curves available at the Mikulski Archive for Space Telescopes\footnote{{\tt {\scriptsize https://mast.stsci.edu/portal/Mashup/Clients/Mast/Portal.html}}} (MAST). This time series spans 24.45\,d and contains 16362 measurements. To determine the pulsation frequencies, we used only the data that were taken out of eclipse, which reduced the data set to 14019 measurements (time span 24.07\,d).

\begin{figure}
\centering
\includegraphics[width=84mm,viewport=35 45 522 470]{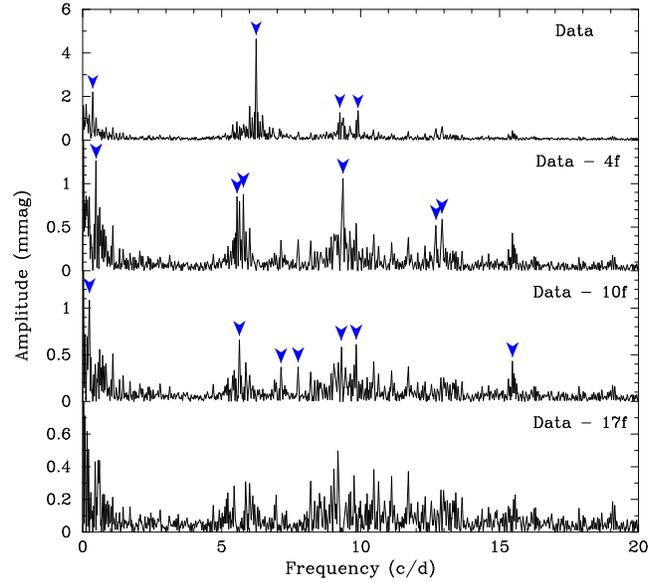}
\caption{The Fourier Transform of the out-of-eclipse \tess\ light curve of AI Hya (top) and subsequent prewhitening steps. The blue arrows denote the signals detected. Outside of the frequency range shown no significant signal is present.}
\label{fig:ft_plot}
\end{figure}

This time series was analysed using the {\sc Period04} software \citep{2005CoAst.146...53L} by group-K. This package applies single-frequency power spectrum analysis and simultaneous multi-frequency sine-wave fitting. These sine-wave fits are subtracted from the data and the residuals examined for the presence of further periodicities. The application of this procedure to AI\,Hya is illustrated in Fig.\,\ref{fig:ft_plot}.

During such a process, it is important to decide where to stop. Often this is facilitated via the application of 
SNR
criteria. In this work, we have adopted the strategy proposed by \cite{1993A&A...271..482B} which is to compute the ratio of the signal amplitude relative to the local noise level to determine whether the frequency under consideration represents a significant detection. Whereas \cite{1993A&A...271..482B} propose 
SNR\,$>$\,4
for a detection, recent findings for space-based data \cite[e.g.,][]{2021AcA....71..113B} suggest that a more conservative limit must be chosen. Given the restricted frequency range in which we search for periodicities, our requirement was 
SNR\,$>$\,4.5.
Furthermore, in unresolved frequency spectra, the periodic content present in the time series can easily be overinterpreted \citep{2014MNRAS.439.3453B} which suggests caution regarding the present data set. Consequently, we stopped the frequency search after the detection of 17 signals (lowest panel of Fig.\,\ref{fig:ft_plot}). More periodicities are certainly present, but these need to await a longer data set for reliable detection. We list the frequency solution so derived in Table\,\ref{tab:puls_freq}.

\begin{table}
\centering
\caption{A least squares fit of the pulsation frequencies of AI Hya. Formal error estimates for the independent frequencies and phases \citep{1999DSSN...13...28M} are given in braces in units of the last digits after the comma.}
\begin{tabular}{lccc}
\hline
&\multicolumn{1}{c}{Frequency} & \multicolumn{1}{c}{Amplitude} &
\multicolumn{1}{c}{SNR}  \\
&\multicolumn{1}{c}{d$^{-1}$} & \multicolumn{1}{c}{mmag} &
\multicolumn{1}{c}{ }   \\
& & \multicolumn{1}{c}{$\pm 0.02$} &
   \\
\hline
$\nu_1$ & 6.2412(1) & 4.75 &  54.2 \\
$\nu_2$ & 9.2654(4) & 1.18 &  9.7 \\
$\nu_3$ & 9.9065(4) & 1.20 &  9.4 \\
$\nu_4$ & 12.715(1) & 0.48 &  4.5 \\
$\nu_5$ & 12.928(1) & 0.54 &  5.4 \\
$\nu_6$ & 9.3689(4) & 1.42 &  11.5 \\
3$\nu_{orb}$ & 0.3619 & 1.76 &  7.5 \\
4$\nu_{orb}$ & 0.4825 & 1.32 &  5.9 \\
$\nu_7$ & 5.5599(7) & 0.78 &  8.3 \\
$\nu_8$ & 5.7804(1) & 0.69 &  7.5 \\
2$\nu_{orb}$ & 0.2413 & 1.75 &  7.3 \\
$\nu_9$ & 5.6375(7) & 0.73 &  7.7 \\
$\nu_{10}$ & 7.136(1) & 0.37 &  6.0 \\
$\nu_{11}$ & 7.751(1) & 0.39 &  5.2 \\
$\nu_{12}$ & 9.3051(6) & 0.82 &  6.7 \\
$\nu_{13}$ & 9.8432(8) & 0.69 &  5.3 \\
$\nu_{3}+\nu_{7}$ & 15.464(1) & 0.43 &  5.6 \\
\hline
\end{tabular}
\label{tab:puls_freq}
\end{table}

This table also contains three harmonics of the orbital period. These are not pulsation frequencies, but a consequence of residual binary-induced variability (see Section on binary modeling for a discussion). The pulsation frequencies themselves were found in an interval between 5.5 -- 13.0\,d$^{-1}$, with one possible combination frequency. It is however not clear whether this is a real combination or just a numerical coincidence keeping in mind the short data set, hence poor frequency resolution. Our frequency solution is similar to that reported by \cite{2020PASJ...72...37L} apart from their identification of possible combination frequencies that are partly implausible.

To use the pulsations to learn more about the individual components by applying asteroseismic methods, it is essential to know from which star the pulsations originate. A quick look at the \tess\ light curve reveals that pulsations are clearly visible during the total part of the primary eclipse, meaning that the secondary is the source of the highest amplitude oscillations. However, both components of AI Hya are located within the pulsational instability strip of the $\delta$ Scuti stars \citep[][see Fig.\,\ref{fig:hr}]{2019MNRAS.485.2380M}, thus the primary may pulsate as well. 
$\delta$ Scuti stars generally pulsate in pressure and mixed modes of low radial order \citep[e.g.,][]{2000ASPC..210....3B}. Using the stellar parameters from Table~\ref{lcresult}, we can compute the expected frequency of the radial fundamental mode of both pulsators from the pulsation constant $Q=P\sqrt{\rho/\rho_{\odot}}=P M^{1/2} R^{-3/2}$, assuming $Q$ to be 0.033\,d for this mode \citep{1981ApJ...249..218F}. We thus expect the radial fundamental mode frequency of the primary component to be around 9.3\,d$^{-1}$, and around 5.8\,d$^{-1}$ for the secondary component, respectively. In Table \ref{tab:puls_freq} oscillation frequencies around both these values are seen, which allows no more than the educated guess that the pulsations below $\sim 8$\,d$^{-1}$ would arise from the secondary component, whereas the higher frequency modes could originate from either star.

A determination of the origin of the pulsations from the orbital light time effect is unfortunately out of reach. The expected light time effect would be about 30\,s (cf. Table\,\ref{rvresult}). An attempt to measure the effect for the strongest pulsation frequency yielded $35 \pm 111$\,s, a null result. To conclude, because it is impossible to say with confidence which pulsation frequencies arise 
from
which component of AI Hya, an asteroseismic analysis cannot be carried out.

\section{Evolutionary models}

The evolutionary status of the binary components was examined by utilizing the Modules for Experiments in Stellar Astrophysics (\textsc{mesa}) evolution code \citep{2011ApJS..192....3P, 2013ApJS..208....4P} which includes a binary module \citep{2015ApJS..220...15P} to examine the binary orbital evolution and to determine the initial parameters of binary systems. In this examination, various evolutionary models were generated considering different metallicity ($Z$). In the models, MESA equation-of-state (EOS) were used. The EOS tables are based on the OPAL EOS tables \citep{2002ApJ...576.1064R}. The OPAL opacity tables and the default solar mixtures were adopted as Z initial fraction from \cite{2009ARA&A..47..481A}. Helium mass fraction were taken Y=0.28, for Z=0.02. Convective core overshoot was described by the exponentially decaying prescription of \cite{2000A&A...360..952H} and overshooting parameter adopted 0.20 for both components (\cite{2016A&A...592A..15C} find 0.208 for both components). A mixing length $\alpha_{MLT}$\, value of 1.8 was used as the theoretical \ds\, instability strip \citep{2004A&A...414L..17D, 2005A&A...435..927D} was obtained with this $\alpha_{MLT}$ value.

\begin{figure}
\centering
\includegraphics[width=8.5cm, angle=0]{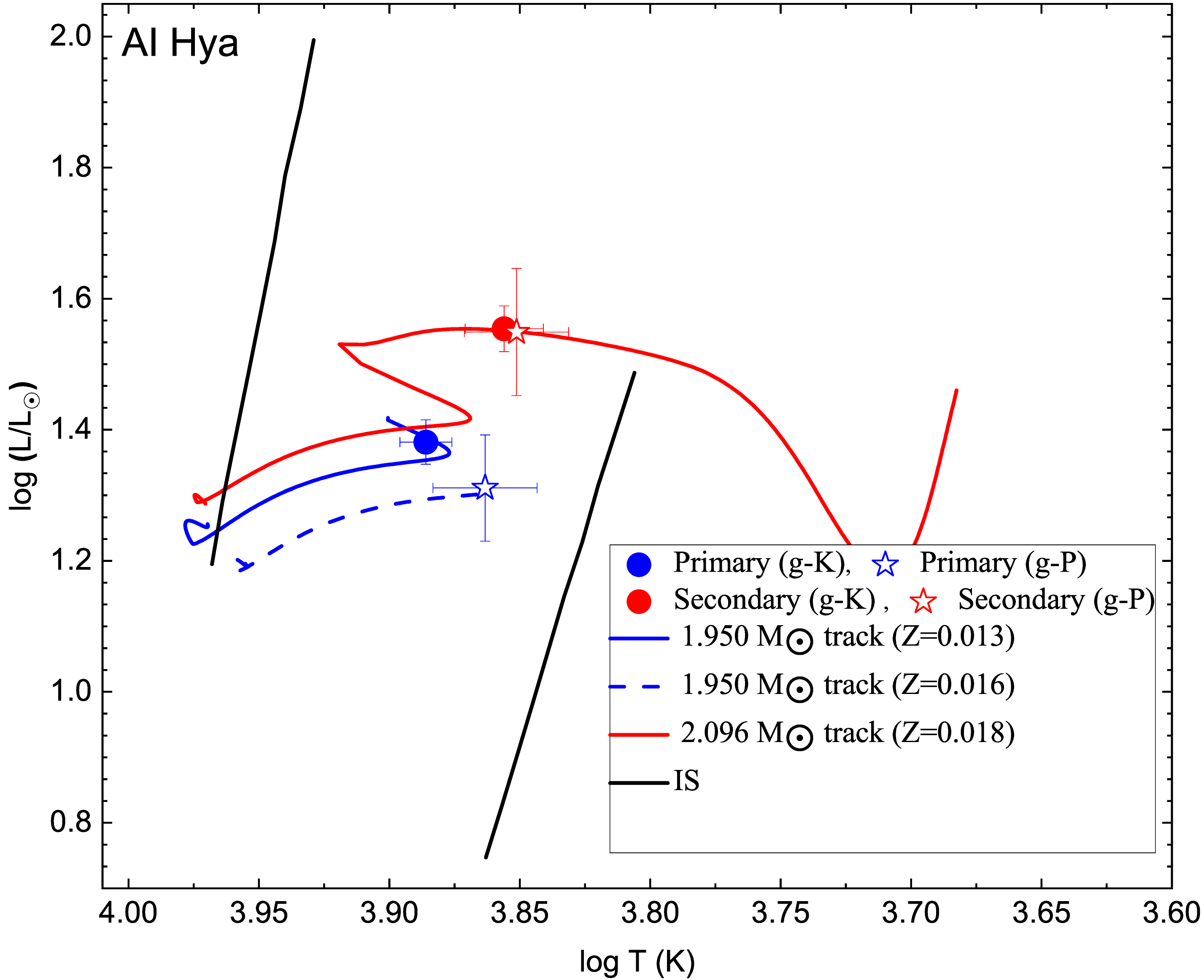}
\caption{The positions of the binary components in the H-R diagram according the results of both group-K (g-K) and group-P (g-P). The instability strip (IS) borders of the \ds\, stars were taken from \citet{2019MNRAS.485.2380M}. }
\label{fig:hr}
\end{figure}

Taking into account the calculated parameters in the binary modelling for both groups, the evolutionary status of the binary components was investigated. As a result, we found that the secondary (more luminous) binary component can be represented with the same evolutionary tracks according to both groups' results. However, the less luminous primary component's position was determined with different $Z$ parameters as the parameters of this star were found to be slightly different in the study of the two groups. According to the evolutionary models, the $Z$ parameters of both binary components were found similar to solar \citep{2009ARA&A..47..481A} within the errors which differs from the results of the groups as we determined that the less luminous component's atmosphere is somewhat enhanced in metals. The results of this analysis are given in Table\,\ref{evoltable} and a H-R diagram is shown in Fig.\,\ref{fig:hr}. The observational borders of the \ds\, instability strip were taken from \cite{2019MNRAS.485.2380M}. As can be seen from the H-R diagram, both binary components are placed inside the \ds\, instability strip.

\section{Discussion and Conclusions}

In this analysis, we present the results of the detailed analysis of AI\,Hya carried out by two independent
groups. The system was observed with different high-resolution spectrographs (R$\gtrsim$38000). The radial velocity variations of AI\,Hya 
were
modelled using the \vr\, measurements of both groups 
and the orbital parameters such as \tzero, \porb, $e$ and $q$ were updated. The resulting parameters of the analysis of both groups are consistent with each other within the errors and they slightly differ from the results of \cite{1988AJ.....95..190P}. Especially the $e$ value shows
a discrepancy. \cite{1988AJ.....95..190P} found $e$ to be 0.2301\,$\pm$\,0.0015
while in our study it was determined as 0.2419\,$\pm$\,0.0036 and 0.2432\,$\pm$\,0.0050 by group-P and -K, respectively. 

Since our high-resolution spectra are spread over all orbital phases, we were able to derive the atmospheric parameters of both binary components by modelling either the composite spectra or the spectra of the individual components after applying spectral disentangling.
To derive the atmospheric parameters, \vsini\, and the chemical composition of the binary components, group-K analysed disentangled spectra of the components, while group-P performed their analysis using both the composite and disentangled spectra. As a result, group-K found that the more luminous star is cooler than the less luminous component.
They found the \teff\, values from the $H_\beta$ line fit and Fe lines to be 7500\,$\pm$\,200\,K and 7700\,$\pm$\,100\,K for the primary and 7000\,$\pm$\,150\,K and 7200\,$\pm$\,100\,K for the secondary component, respectively. Group-P used two different codes in their analysis. With the \textsc{gssp} code analysis they found a similar result with group-K even though the resulting \teff\, values differ from each other, they determined that the more luminous star is cooler (7150\,$\pm$\,250\,K) and less luminous one is hotter (7350\,$\pm$\,300\,K). In the \textsc{iSpec} analysis of group-P, \teff\, values of both components were found similar to the results of the \textsc{gssp} analysis within error bars. The primary's temperature is the most significant discrepancy between the values derived by the two groups.
The exact reason for this temperature inconsistency is not fully understood, although it is still only at a level of $\sim$1.1$\sigma$.

In the chemical abundance analysis, both groups found the less luminous but hotter binary component to
show overabundance while the other component has chemical abundance similar to solar. Both groups determined the abundances of some individual elements such as iron (Fe). They derived Fe abundances as 8.25\,$\pm$\,0.23 (group-K) and 7.83\,$\pm$\,0.16 (group-P). These values are consistent with each other within their 1$\sigma$ errors, and both demonstrate that the hotter component has a slightly metal-rich chemical abundance compared to solar values 
(see Table\,\ref{abunresult}). This comes somewhat to a surprise, as this binary system should have been formed in the same interstellar environment and hence its components should have the same chemical composition. 
The difference could be due to the consequences of the evolution of the system. If AI Hya had a very eccentric orbit when the system
was
formed, there could be some material flows from one component to another that could have changed the diffusion in one component. Another explanation was given by \cite{2015AJ....149...59Y} and they pointed out that possible gas and dust accretion from the circumstellar envelope could alter the atmospheric composition of one component. 

After the determination of the atmospheric parameters, they were used as input in the binary modelling. Overall, even though both working groups used different approaches to estimate the parameters of the binary component of AI\,Hya, 
the values determined by both groups are found to be consistent with each other within the error bars. 
The two groups obtained very similar $M$ and $R$ values 
with a 
$\leq$1.7\% and $\sim$0.5\% accuracy, respectively. When we compare these values with the ones found by \cite{2020PASJ...72...37L}, we notice that there are slight differences, especially in the $R$ parameters, and there is significant diversity in the calculated distance. These differences could be caused by the different assumptions of the atmospheric parameters.

The evolutionary status of the system was examined and it was found that both binary components are inside the \ds\, instability strip. The age of the system is determined as well. According to the determined ages, we could say that AI\,Hya is in an important evolutionary phase in terms of binary evolution. The rapidly evolving massive component will begin the mass transfer process to the less massive one approximately 20\,Myr from now. This situation could cause significant variations in the oscillation properties. Increasing the number of such bodies is important in terms of examining the pulsating structures before the mass transfer processes.   

\begin{table}
\centering
\caption{Results obtained from the best-fit evolutionary models.}
\begin{tabular*}{0.95\linewidth}{@{\extracolsep{\fill}}lcc}
\hline
  Parameter                & Group-K      & Group-P   \\
                           &              &            \\
\hline
$P$$_{initial}$  (days)    & 8.34 (1)     & 8.34 (1)   \\
$e$$_{initial}$            & 0.242 (2)    & 0.243 (2)  \\
$Z$$_{1}$                  & 0.013 (2)    & 0.016 (2)  \\
$Z$$_{2}$                  & 0.018 (2)    & 0.018 (2)  \\
Age (Myr)                  &850 (20)     & 860 (20)   \\
\hline
\end{tabular*}
\label{evoltable}
\end{table}

The pulsation properties of AI\,Hya were examined using the \tess\ data. However, the system has only one sector of SC 
data,
which offers us a poor frequency resolution. In the analysis, pulsation frequencies were found between 5.5 and 13\,d$^{-1}$. As both binary components are placed in the 
\ds\ instability strip,
we were unable to say whether one or both pulsate. Apart from that, we could not find pulsations related to the orbital frequency. 

As a result of this study, we thoroughly examined a detached binary system showing oscillations. This kind of objects is particularly important to examine the instability strip of $\delta$\,Scuti stars since they allow us to determine fundamental astrophysical, atmospheric parameters and the chemical abundances of individual binary components. Hence an increasing number of analyses of such systems is expected to be essential to deeply understand the nature of pulsations.  

\section*{Acknowledgments}
The authors would like to thank the reviewer for useful comments and suggestions that helped to improve the publication. This study  has  been  supported by  the  Scientific  and  Technological  Research  Council (TUBITAK) project 120F330. GH thanks the Polish National Center for Science (NCN) for supporting the study through grants 2015/18/A/ST9/00578 and 2021/43/B/ST9/02972. TP's research is supported through NCN OPUS project number 2017/27/B/ST9/02727. AM's acknowledges the support provided by the Polish National Science Centre (NCN) OPUS project number 2017/27/B/ST9/02727 and 2021/41/N/ST9/02746. Based  on observations  made  with  the Mercator  Telescope,  operated on  the  island  of  La  Palma  by  the Flemish Community, at the Spanish Observatorio del Roque de los Muchachos of the Instituto de Astrof\`{\i}sica de Canarias. The \tess\ data presented in this paper were obtained from the Mikulski Archive for Space Telescopes (MAST). Funding for the \tess\ mission is provided by the NASA Explorer Program. This work has made use of data from the European Space Agency (ESA) mission Gaia (http://www.cosmos.esa.int/gaia), processed by the Gaia Data Processing and Analysis Consortium (DPAC, http://www.cosmos.esa.int/web/gaia/dpac/consortium). Funding for the DPAC has been provided by national institutions, in particular the institutions participating in the Gaia Multilateral Agreement. This research has made use of the SIMBAD data base, operated at CDS, Strasbourq, France. 

\section*{DATA AVAILABILITY}
The data underlying this work will be shared at reasonable request to the corresponding author.

\newpage

\appendix
\setcounter{table}{0}
\renewcommand{\thetable}{A\arabic{table}}

\begin{table}
\begin{center}
\caption[]{The $v_{r}$ measurements. The subscripts ``1'' and ``2'' 
  represent the more and the less luminous components, respectively.}\label{rvm}
 \begin{tabular}{lccc}
  \hline\noalign{\smallskip}
HJD         &  $v_{r,1}$       &  $v_{r,2}$   &Instrument           \\
+2450000     &  (km\,s$^{-1}$)              &   (km\,s$^{-1}$)         \\
\hline\noalign{\smallskip}
9263.45270   & -12.6\,$\pm$\,2.8 & 109.3\,$\pm$\,2.7 & CAOS \\
\hline
9161.65803   & 132.8\,$\pm$\,1.8  & -48.3\,$\pm$\,1.7 & HERMES \\
9162.64306   & 109.8\,$\pm$\,2.0  & -21.7\,$\pm$\,1.5 & HERMES \\
9230.65226   & 121.5\,$\pm$\,1.6  & -24.7\,$\pm$\,1.8 & HERMES \\
9231.66393   & 124.1\,$\pm$\,1.7  & -24.9\,$\pm$\,1.7 & HERMES \\
9233.62648   & 59.8\,$\pm$\,5.7   & 36.1\,$\pm$\,3.4  & HERMES \\
9234.55784   & 20.6\,$\pm$\,2.1   & 72.1\,$\pm$\,2.5  & HERMES \\
9237.61273   & 15.0\,$\pm$\,1.6   & 77.3\,$\pm$\,1.5  & HERMES \\
9235.43315   &-46.3\,$\pm$\,1.5   & 130.3\,$\pm$\,1.8 & HERMES \\
9257.49195   & 98.8\,$\pm$\,1.8   & -2.9\,$\pm$\,2.0  & HERMES \\
9260.61123   &-33.9\,$\pm$\,1.7   &117.9\,$\pm$\,1.9  & HERMES \\
9276.55613   &96.6\,$\pm$\,2.0    & -8.4\,$\pm$\,1.2  & HERMES \\
9296.42427   & 88.7\,$\pm$\,1.7   & -3.8\,$\pm$\,2.0  & HERMES \\
9297.44747   & 126.7\,$\pm$\,1.8  & -37.4\,$\pm$\,2.2  & HERMES \\
9298.45846   &113.5\,$\pm$\,1.7   & -16.0\,$\pm$\,1.8 & HERMES \\
9299.46357   & 78.1\,$\pm$\,2.1   & 12.5\,$\pm$\,2.3  & HERMES \\
\hline
7075.62231   & -39.7\,$\pm$\,1.5 & 129.9\,$\pm$\,0.5 & CORALIE \\
7076.63954   & -20.4\,$\pm$\,1.2 & 120.0\,$\pm$\,1.3 & CORALIE \\
7109.63123   & -17.9\,$\pm$\,2.4 & 126.0\,$\pm$\,1.3 & CORALIE \\
\hline
7022.31643   & 118.5\,$\pm$\,1.9 & -31.4\,$\pm$\,0.5 & HIDES \\
7060.09414   & -13.6\,$\pm$\,0.7 & 118.2\,$\pm$\,0.6 & HIDES \\
7109.96513   & -18.6\,$\pm$\,1.1 & 114.6\,$\pm$\,0.6 & HIDES \\
7114.92732   & 120.1\,$\pm$\,1.2 & -34.0\,$\pm$\,0.7 & HIDES \\
7146.98403   & 118.5\,$\pm$\,1.3 & -42.9\,$\pm$\,0.8 & HIDES \\
7147.96084   & 131.3\,$\pm$\,1.2 &-42.3\,$\pm$\,0.6  & HIDES \\
7363.28986   & 135.7\,$\pm$\,0.6 &-48.4\,$\pm$\,0.7  & HIDES \\
7755.22744   &-34.0\,$\pm$\,0.7  &126.5\,$\pm$\,0.7  & HIDES \\
7813.13416   & -25.2\,$\pm$\,0.8 &124.3\,$\pm$\,0.9  & HIDES \\
7814.08321   & -28.3\,$\pm$\,0.9 &126.6\,$\pm$\,0.9  & HIDES \\
7846.01822   & -10.8\,$\pm$\,1.7 &113.4\,$\pm$\,1.0  & HIDES \\
8035.34461   & 101.6\,$\pm$\,0.7 &-13.9\,$\pm$\,0.8  & HIDES \\
8066.24908   & 85.2\,$\pm$\,0.8  & -4.1\,$\pm$\,1.3  & HIDES \\
  \noalign{\smallskip}\hline
\end{tabular}
\end{center}
\end{table}

\bsp	
\label{lastpage}
\end{document}